\documentclass[12pt,epsf]{article}
\usepackage{verbatim}
\usepackage{anyfontsize}
\usepackage{dsfont}
\usepackage{tikz}
\usepackage{sidecap}
\sidecaptionvpos{figure}{t}
\usepackage{subfigure}
\usepackage{enumitem}
\usepackage[english]{babel}
\usepackage{enumitem}  

\usepackage{amssymb,amsmath}
\usepackage{graphicx, xcolor, varwidth}
\usepackage{setspace}
\usepackage[permil]{overpic}
\usepackage{cite}
\usepackage{mathtools}

\DeclareSymbolFont{matha}{OML}{txmi}{m}{it}
\DeclareMathSymbol{v}{\mathord}{matha}{118}

\usepackage[framemethod=default]{mdframed}
\newmdenv[skipabove=7pt,
skipbelow=7pt,
rightline=false,
leftline=false,
topline=false,
bottomline=false,
backgroundcolor=blue!10,
linecolor=blue,
innerleftmargin=5pt,
innerrightmargin=5pt,
innertopmargin=5pt,
innerbottommargin=5pt,
leftmargin=0cm,
rightmargin=0cm,
linewidth=4pt]{bBox}

\colorlet{darkblue}{blue!70!black}
\colorlet{darkgreen}{green!70!black}

\usepackage[colorlinks=true,urlcolor=darkblue,linktocpage=true,linkcolor=darkblue,citecolor=darkblue]{hyperref}
\hypersetup{ 
colorlinks=true, 
linkcolor=blue, 
citecolor=magenta, 
}

\numberwithin{equation}{section}
\DeclareMathSymbol{v}{\mathord}{matha}{118}
\newcommand{\be}{\begin{equation}}
\newcommand{\ee}{\end{equation}}
\newcommand{\bea}{\begin{eqnarray}}
\newcommand{\eea}{\end{eqnarray}}
\newcommand{\bear}{\begin{eqnarray}}
\newcommand{\eear}{\end{eqnarray}}
\newcommand{\beas}{\begin{eqnarray*}}
\newcommand{\p}{\partial}
\newcommand{\eeas}{\end{eqnarray*}}
\newcommand{\ba}{\begin{array}}
\newcommand{\ea}{\end{array}}

\def\ba#1\ea{\begin{align}#1\end{align}}
\def\bs#1\es{\begin{split}#1\end{split}}

\newcommand{\tr}{\operatorname{tr}}
\newcommand{\pd}[2][1]{\ifnum#1=1 \frac{\partial}{\partial {#2}} \else
  \frac{\partial^#1}{\partial {#2}^{#1}}\fi}
\newcommand{\dpd}[2][1]{\ifnum#1=1 \dfrac{\partial}{\partial {#2}} \else
  \frac{\partial^#1}{\partial {#2}^{#1}}\fi}
\newcommand{\td}[2][1]{\ifnum#1=1 \frac{d}{d{#2}} \else
  \frac{d^#1}{d{#2}^{#1}}\fi}

\renewcommand{\(}{\left(}
\renewcommand{\)}{\right)}

\renewcommand{\]}{\right]}

\newcommand{\nbox}{{\,\lower0.9pt\vbox{\hrule \hbox{\vrule height 0.2 cm \hskip 0.19 cm \vrule height 0.2 cm}\hrule}\,}}

\newcommand{\ie}{{\it i.e.,}\ }

\def\O{{\cal O}}
\renewcommand{\S}{{\mathcal S}}

\newcommand{\N}{{\cal N}}



\begin{document}
\begin{spacing}{1.3}
\begin{titlepage}

\begin{center}
{\Large 
\vspace*{6mm}

\bf Extremal Chaos

}

\vspace*{6mm}

Sandipan Kundu

\vspace*{6mm}

\textit{Department of Physics and Astronomy,
\\ Johns Hopkins University,
Baltimore, Maryland, USA\\}

\vspace{6mm}

{\tt \small kundu@jhu.edu}

\vspace*{6mm}
\end{center}

\begin{abstract}
In maximally chaotic quantum systems, a class of out-of-time-order correlators (OTOCs) saturate the Maldacena-Shenker-Stanford (MSS) bound on chaos. Recently, it has been shown that the same OTOCs must also obey an infinite set of (subleading) constraints in any thermal quantum system with a large number of degrees of freedom. In this paper, we find a unique analytic extension of the maximally chaotic OTOC that saturates all the subleading chaos bounds which allow saturation. This extremally chaotic OTOC has the feature that information of the initial perturbation is recovered at very late times. Furthermore, we argue that the extremally chaotic OTOC provides a K\"{a}llen-Lehmann-type representation for all OTOCs. This representation enables the identification of all analytic completions of maximal chaos as small deformations of extremal chaos in a precise way.

\end{abstract}

\end{titlepage}
\end{spacing}

\vskip 1cm
\setcounter{tocdepth}{2}  
\tableofcontents

\begin{spacing}{1.3}


\section{Introduction}
In recent years, it has been exceedingly clear that theories of quantum gravity and their holographic duals are {\it maximally chaotic}. This idea leverages the striking insight that there is a universal upper bound on quantum chaos \cite{Maldacena:2015waa}. The Maldacena-Shenker-Stanford (MSS) bound  \cite{Maldacena:2015waa} states that  in thermal quantum systems with a large number of degrees of freedom, a class of out-of-time-order correlators (OTOCs), which measures chaos,  cannot grow with time faster than $e^{\lambda_L t}$, with a Lyapunov exponent $\lambda_L\le 2\pi/\beta$. This observation provides a precise definition of the general idea of maximal chaos \cite{Sekino:2008he} in terms of OTOCs that saturate the MSS bound. However, it is also known that this description is incomplete since these OTOCs are also bounded and hence cannot grow indefinitely. This conflict implies that the rate of growth of all maximally chaotic OTOCs must deviate significantly from the MSS saturation as they approach the {\it scrambling time} $t_*$. In other words, maximal chaos is a statement about the leading order behavior of an OTOC, which requires an {\it analytic completion}.  So, an important problem is to systematically study analytic completions of maximal chaos, finding universal features. One of the goals of this paper is to develop some tools to address this problem. 

Interestingly, the above problem is closely related to another more conceptual problem that has emerged from a new set of chaos bounds obtained in \cite{Kundu:2021qcx}. In particular, it was shown in \cite{Kundu:2021qcx} that the same class of OTOCs must satisfy an infinite set of constraints. The MSS bound, which is just one of these constraints, can be regarded as the leading chaos bound. However, there are infinitely many additional  {\it subleading chaos bounds} that, in principle, can also be saturated. So, what could be more natural than to ask whether, and in what way, all these chaos bounds can be consistently saturated? It is almost unsurprising that there is a connection between this problem and the problem of analytic completions of maximal chaos.

Let us now make these questions more precise. The MSS bound applies to all unitary chaotic systems with a large number of degrees of freedom and a simple Hamiltonian $H$. In such a system, consider the thermal OTOC at temperature $T=1/\beta$
\be\label{eq:otoc}
F(t)=\tr \[y V(0) y W(t)yV(0)yW(t)\]\ , \qquad y^4=\frac{e^{-\beta H}}{\tr\[e^{-\beta H}\]}\ 
\ee
of any two simple Hermitian local operators $V$ and $W$ with vanishing thermal one-point functions. This OTOC  is an  analytic function  in the half-strip $\{t\in \mathbb{C}|\ \mbox{Re}\ t>0\ \text{and}\ |\mbox{Im}\ t|\le \frac{\beta}{4} \}$, obeying the Schwarz reflection condition $\(F(t)\)^*=F\(t^*\)$. The MSS bound was derived under an additional well-motivated assumption that this OTOC is also bounded $|F(t)|\le F_d$  on the boundary $\ |\mbox{Im}\ t|= \frac{\beta}{4}$ of the half-strip by the factorized correlator\footnote{For a  definition see equation (\ref{def:Fd}).} above the {\it factorization time scale} $\mbox{Re}\ t \ge t_0$, where  the factorization time $t_0$  is larger than the dissipation time but much smaller than the scrambling time $t_*$. This imposes a sharp bound on the rate of growth $ \( \frac{2\pi}{\beta}- \p_t\)\(F_d-F(t)\)\ge 0$ for $t\gg t_0$ \cite{Maldacena:2015waa}.\footnote{Originally, the MSS bound was derived in \cite{Maldacena:2015waa} using a stronger boundedness assumption. It was argued in  \cite{Kundu:2021qcx} that the stronger assumption is not necessary for the proof of the MSS bound. }

A quantum system is said to be maximally chaotic when OTOCs exhibit a period of Lyapunov growth saturating the MSS bound at the leading order 
\be\label{intro:maximal}
F(t)=F_d-\frac{c_1}{\N} e^{\frac{2\pi}{\beta} t}+\cdots \ , \qquad \text{for}\qquad t_0\ll t\ll t_*\ ,
\ee 
where $c_1$ is a positive order one coefficient and $\N\gg 1$ is an effective measure of the number of degrees of freedom per site, determining the scrambling time $t_*=\frac{1}{2\pi}\ln \N$ \cite{Sekino:2008he}. However, the maximally chaotic OTOC (\ref{intro:maximal}), by itself, is neither bounded nor analytic in the entire  half-strip $\S=\{t\in \mathbb{C}|\ \mbox{Re}\ t\ge t_0\ \text{and}\ |\mbox{Im}\ t|\le \frac{\beta}{4} \}$. Hence, the maximally chaotic OTOC must be accompanied by correction terms that make it consistent with the analyticity and the boundedness conditions in the entire half-strip $\S$. These correction terms, which are denoted by dots in (\ref{intro:maximal}), start to dominate at some time scale $t_{\rm eff}\le t_*$.  In other words,  the full OTOC $F(t)$ is analytic and bounded   in $\S$, approaching the maximally chaotic OTOC (\ref{intro:maximal}) only at early times $t<t_{\rm eff}$. We can now ask the following question. What rigorous statements   can we make about the full OTOC from basic principles and symmetries? This problem is worth exploring since this class of OTOCs will be of importance in quantum gravity.

Recently, it was shown in \cite{Kundu:2021qcx} that the OTOC (\ref{eq:otoc}), under the same set of assumptions, must satisfy an infinite set of additional local constraints beyond the MSS bound. These new chaos bounds also constrain the correction terms in (\ref{intro:maximal}). In itself, this should not be too surprising since not any early-time expansion of $F(t)$ can resum into a function that is analytic and bounded even at late times. The chaos bounds of \cite{Kundu:2021qcx} provide a systematic  realization of this fact by introducing a local {\it moment} $\mu_J(t)$ of the OTOC, as defined in (\ref{def:moments}). These bounds state that the moment $\mu_J(t)$, for integer $J\ge 0$, must be  a positive, bounded, monotonically decreasing, log-convex function of $J$ for all real $t\ge t_0$ \cite{Kundu:2021qcx}.\footnote{An analogous but strictly weaker statement is that $\left[\prod_{I=1}^N \( \frac{2\pi(2I-1)}{\beta}- \p_t\)\right]\(F_d-F(t)\)\ge 0$ for all integer $N\ge 1$ and $t\gg t_0$ \cite{Kundu:2021qcx}. Of course, the MSS bound is the special case $N=1$.} Importantly, this includes an infinite subset of bounds that allow saturation. So, from the perspective of these new bounds, the maximally chaotic OTOC (\ref{intro:maximal}) appears to be conceptually incomplete since it saturates only one of the infinitely many constraints. This leads to the idea of extremal chaos, which we explain next. 

\begin{figure}
\centering
\includegraphics[scale=0.3998]{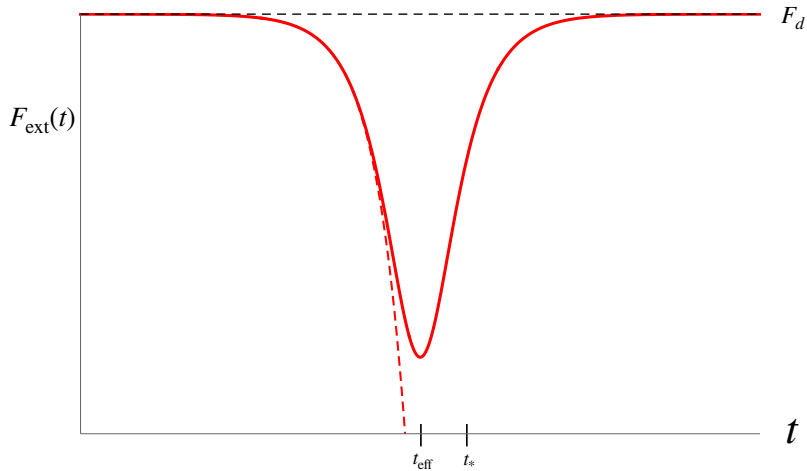}
\caption{ \label{figure:intro} \small The extremally chaotic OTOC $F_{\rm ext}(t)$ coincides with the maximally chaotic OTOC (red dashed line) only before the effective time scale $t_{\rm eff}$. The extremally chaotic OTOC has a minimum at $t=t_{\rm eff}$, irrespective of the scrambling time $t_*$. After $t=t_{\rm eff}$, the OTOC grows monotonically approaching the factorized value (also the initial value) $F_d$. So, $F_{\rm ext}(t)$ is bounded by the factorized correlator $F_d$ for real $t$ even after the scrambling time. }
\end{figure}

At this stage, we have compelling reasons to ask whether there are OTOCs that satisfy all of the following conditions:
\begin{enumerate}
\item{$F(t)$ saturates, in a consistent way, all the chaos bounds of \cite{Kundu:2021qcx} that allow saturation. }
\item{$F(t)$ coincides with the maximally chaotic OTOC (\ref{intro:maximal}) at early times. }
\item{$F(t)$, as a function of complex $t$, is analytic inside the entire half-strip $\{t\in \mathbb{C}|\ \mbox{Re}\ t\ge t_0\ \text{and}\ |\mbox{Im}\ t|< \frac{\beta}{4} \}$.}\footnote{Note that we are excluding the boundary of the half-strip $|\mbox{Im}\ t|=\frac{\beta}{4}$. There is no non-trivial solution of this set of conditions when the boundary is included. This fact has important implications, as we will discuss later.}
\end{enumerate}
At first sight, the above conditions seem to be overconstraining. So, it is rather surprising that there is a unique solution to the above set of conditions 
\be\label{eq:intro}
F_{\rm ext}(t)= F_d -   \frac{c_1}{\N} \mathcal{F}_{\rm ext}(t;t_{\rm eff})\ , \qquad  \mathcal{F}_{\rm ext}(t;t_{\rm eff})=\frac{e^{\frac{2\pi}{\beta}t}}{1+ e^{\frac{4 \pi}{\beta} (t-t_{\rm eff})}}
\ee
up to terms that decay exponentially for $t\gg t_0$. We will refer to this solution as the {\it extremally chaotic} (or {\it extremal}) OTOC.\footnote{Note that extremal chaos, as defined in this paper, should not be confused with chaos in extremal black holes, which is discussed in \cite{Poojary:2018esz,Banerjee:2019vff,Craps:2020ahu,Craps:2021bmz}.} In the above result,  $t_{\rm eff}$ is an  {\it effective} time scale at which correction terms in (\ref{intro:maximal}) become significantly large. The extremally chaotic OTOC reaches its minimum value at $t=t_{\rm eff}$ indicating maximal scrambling of the initial perturbation.  So, in some sense, $t_{\rm eff}$ is the effective scrambling time. However, in general $t_{\rm eff}$ is independent of the traditional scrambling time $t_*$.  

The extremally chaotic OTOC grows monotonically for $t>t_{\rm eff}$, as shown in figure \ref{figure:intro}. So, the information of the initial perturbation is not completely lost. In particular,  this information can be fully recovered in the limit $t\rightarrow \infty$ at which $F_{\rm ext}(t)\rightarrow F_d$. In contrast, thermal OTOCs of large $N$ holographic CFTs asymptote to zero for $t>t_*$, a fact that can be deduced  for heavy operators from the large $c$ identity Virasoro block in 2d CFT \cite{Roberts:2014ifa} and from the elastic eikonal approximation in 3d gravity \cite{Shenker:2014cwa}.

Interpretation of the extremal solution (\ref{eq:intro}), however, is more subtle. This is because the extremal OTOC has singularities at $t=t_{\rm eff}\pm i \frac{\beta}{4}$. From this non-analyticity, one could conclude that the extremal solution (\ref{eq:intro}) is unphysical, but this would be premature. The non-analyticity simply means that $F_{\rm ext}(t)$ should be interpreted as a distribution.\footnote{A closely related general statement is the Vladimirov’s theorem \cite{Vladimirov} (for a recent review see \cite{Kravchuk:2020scc}).} To give this statement a definite meaning, we next introduce a spectral representation of the OTOC (\ref{eq:otoc}).

In this paper, we will also argue that there exists  a K\"{a}llen-Lehmann-like representation of the OTOC (\ref{eq:otoc}): 
\be\label{intro:KL}
 F_d-F(t)=\int_{t_0}^\infty dt' \mathcal{F}_{\rm ext}(t;t')\rho(t')\ , \qquad 0\le \rho(t')\le \frac{8}{\beta}e^{-\frac{2\pi t'}{\beta}}F_d
 \ee
for $\mbox{Re}\ t\gg t_0$ and $|\mbox{Im}\ t|< \beta/4$, where $\mathcal{F}_{\rm ext}(t;t')$ is the extremal OTOC (\ref{eq:intro}). This is true even for OTOCs that are not maximally chaotic in any duration of time. Hence, $\mathcal{F}_{\rm ext}(t;t')$ has a natural interpretation as a universal distribution,\footnote{We cannot help but notice that the distribution $\mathcal{F}_{\rm ext}(t;t')$, as a function of $t'$, looks very similar to the Fermi-Dirac distribution, which is perhaps just a coincidence. Nevertheless, this observation enables us to utilize various mathematical tools available for the Fermi gas to analyze quantum chaos. } whereas $\rho(t')$ can be thought of as a theory-dependent {\it density function}.  It is possible to write an inversion formula for the density function $\rho(t')=\frac{4}{\beta}e^{-\frac{2\pi t'}{\beta}}\(F_d-\mbox{Re}\ F\(t'+i\frac{\beta}{4}\)\)$ that implies the two-sided bound for the density function in (\ref{intro:KL}).\footnote{The inversion formula also implies that the density function $\rho(t')$ is a smooth function of class $C^\infty$.} 

The representation (\ref{intro:KL}) has a significant technical as well as conceptual advantage. Any OTOC written in the form (\ref{intro:KL}) is automatically consistent with all the chaos bounds. So, the representation (\ref{intro:KL}) provides a natural language to study the OTOC (\ref{eq:otoc}) in physical systems with many degrees of freedom. This framework is going to be particularly useful for studying analytic completions of maximal chaos.

\begin{figure}[h]
\centering
\includegraphics[scale=0.51]{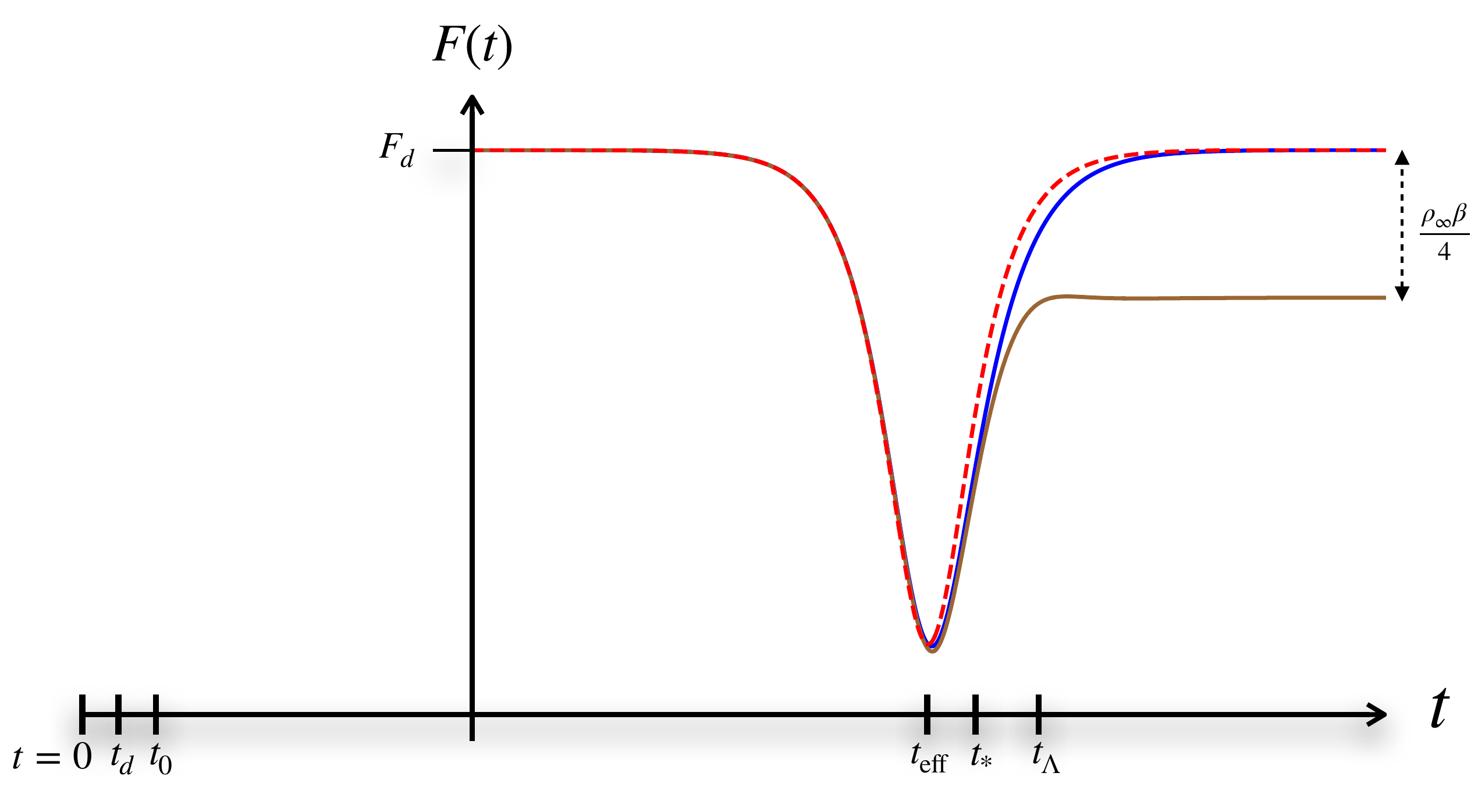}
\caption{ \label{figure:intro2} \small  Extremal chaos in physical systems. The extremally chaotic OTOC  (dashed red) is non-analytic on the boundary: $\mbox{Im}\ t=\pm \beta/4$. This non-analyticity can be removed by a standard $i\epsilon$-shift. The associated regularized OTOC, which is shown in blue, differs only slightly from the unregularized OTOC.  However, a late-time long-tailed correction to the density function $\rho(t>t_\Lambda)=\rho_\infty e^{-\frac{2 \pi }{\beta}t} $ does change the asymptotic behavior of the OTOC (shown in brown). Various relevant time scales are also shown in the figure: $t_d$= dissipation time $\sim \beta$, $t_0$= factorization time, $t_{\rm eff}=$ effective (or a gap) time scale, $t_*=$ scrambling time, and $t_\Lambda>t_*$ is a cut-off scale above which late-time corrections can become important.  In the presence of a late-time long-tailed correction, a physical system can be extremally chaotic  only approximately up to the cut-off scale $t_\Lambda$.  }
\end{figure}

The extremal OTOC (\ref{eq:intro}) is well-behaved even after the scrambling time, however, it is not a true analytic completion of maximal chaos.  The extremal OTOC is characterized by a density function  $\rho(t')=\frac{c_1}{\N}\delta(t'-t_{\rm eff})$ that is at odds with the boundedness property (\ref{intro:KL}). This tension is a manifestation of the fact that the extremal OTOC has singularities at $t=t_{\rm eff}\pm i \frac{\beta}{4}$. Fortunately, these kinds of singularities are very familiar to us from quantum field theory (QFT). We adopt the standard  $i\epsilon$-prescription of QFT and move these singularities outside the half-strip $\{t\in \mathbb{C}|\ \mbox{Re}\ t>0\ \text{and}\ |\mbox{Im}\ t|\le \frac{\beta}{4} \}$. This $i\epsilon$-regularization essentially replaces the delta function in $\rho(t')$ by a narrow distribution obeying  the boundedness property (\ref{intro:KL}). The resulting regularized extremal OTOC can be computed exactly by using the representation (\ref{intro:KL}) even when $\epsilon$ is finite. The regularized OTOC differs only slightly from the extremal OTOC (\ref{eq:intro}) for real $t$, as shown in figure \ref{figure:intro2}.  So in principle, a physical system, to a good approximation, can be extremally chaotic. 

The $i\epsilon$-regularized extremal OTOC is a ``tree-level" analytic completion of maximal chaos (\ref{intro:maximal}), which asymptotes to the extremally chaotic OTOC  (\ref{eq:intro}) away from the time scale $t_{\rm eff}$.\footnote{In theories of quantum gravity, the scrambling time is determined by the Newton constant $G_N$. The time scale $t_{\rm eff}$ can be thought of as the analog of the string scale.} So, it is expected that the density function, in general, will have higher-order $1/\N$ corrections. The extremal OTOC, regularized or unregularized, has the property that its general structure is insensitive to almost all of these corrections. Only a late-time {\it long-tailed correction}\footnote{Note that the density function cannot decay slower than $e^{-\frac{2 \pi }{\beta}t}$ because of the boundedness condition (\ref{intro:KL}).} to the density function $\delta\rho(t>t_\Lambda)=\rho_\infty e^{-\frac{2 \pi }{\beta}t}$, where $t_\Lambda>t_*$ is a cut-off scale, can affect the OTOC significantly by changing the asymptotic behavior as $t\gg t_\Lambda$. Interestingly, the resulting OTOCs typically have a familiar shape with a late-time plateau, as shown in  figure \ref{figure:intro2}. We expect that higher-order $1/\N$ effects will generate such long-tailed corrections to the density function. This expectation is indeed realized in the Schwarzian theory, as we will show in appendix \ref{app:ST}.

Finally, we come back to the question of general analytic completions of the maximally chaotic OTOC. In general,  the OTOC (\ref{intro:maximal}) can have complicated analytic completions that differ significantly from the extremal OTOC. Nevertheless, we will argue that all analytic completions of a long period of maximal chaos, in the representation (\ref{intro:KL}), must be small deformations of extremal chaos. More precisely, the associated density function is a narrow distribution such that the integral $\int_{t_0}^\infty dt' \rho(t')$ is dominated by a small region $t_{\rm eff}-\Delta t_{\rm eff} \le t \le t_{\rm eff} +\Delta t_{\rm eff}$, where $\Delta t_{\rm eff} \ll t_{\rm eff}\lesssim t_*$. This observation enables a systematic analysis of OTOCs that analytically complete the maximally chaotic OTOC both at early and late times.

There is compelling evidence suggestive of the fact that chaos has a hydrodynamic origin in maximally chaotic systems \cite{Blake:2017ris, Blake:2021wqj}. It would be interesting to study the results of this paper in the hydrodynamic effective field theory (EFT) framework of \cite{Blake:2017ris,Blake:2021wqj}. In particular, the EFT approach might be a good guide in further understanding what extremal chaos physically means. There is also a related interesting story of pole-skipping \cite{Blake:2017ris,Grozdanov:2017ajz,Haehl:2018izb,Blake:2018leo,Grozdanov:2018kkt,Haehl:2019eae,Ahn:2019rnq,Ahn:2020bks,Ramirez:2020qer,Choi:2020tdj} as a way of understanding maximal chaos, which may provide additional insights.

The rest of the paper is organized as follows. We begin with a review of the chaos bounds in section \ref{sec:chaos}. In section \ref{sec:extremal}, we derive the extremally chaotic OTOC (\ref{eq:intro}) and argue for its uniqueness. In section \ref{sec:KL}, we introduce the spectral representation (\ref{intro:KL}) and use it to regularize the extremally chaotic OTOC. Besides, we also study various corrections to extremal chaos. In section \ref{sec:maximal}, we argue that general analytic completions of maximal chaos are small deformations of extremal chaos and discuss its implications. Finally, we end with concluding remarks in section \ref{sec:conclusions}.

\section{Bounds on Chaos}\label{sec:chaos}
In this section, we review some general properties of chaos for thermal quantum systems with a large number of degrees of freedom. In such a system, consider any two simple Hermitian local operators $V$ and $W$ with vanishing thermal one-point functions. The OTOC (\ref{eq:otoc}) measures the effect of a small perturbation induced by the operator $V$ on another operator $W$ at a later time $t>0$. In recent years, this thermal OTOC has emerged as a good measure of quantum chaos since it has some rather nice features. First of all, the OTOC (\ref{eq:otoc}) is  thermally regularized and hence it does not require additional regularization  in quantum field theory to remove coincident point singularities. Secondly, this OTOC enjoys certain analyticity and boundedness properties providing us with some level of mathematical control.

In this paper, we are interested in interacting unitary quantum systems with a large number of degrees of freedom in which the Hamiltonian is made out of finite products of simple operators. In such systems, for times larger than the dissipation time (also known as the local thermalization time) $t_d\sim \beta$ but before the onset of chaos, the OTOC (\ref{eq:otoc}) can be well approximated by the factorized correlator 
\be\label{def:Fd}
F_d= \tr\[y^2 V(0) y^2 V(0)\]\tr\[y^2 W(t)y^2W(t)\]>0\ ,
\ee
irrespective of the choice of operators. If the system is chaotic, the OTOC $F(t)$ starts decreasing rapidly for $t\gg t_d$. This leads to a new time scale relevant for quantum chaos, the {\it scrambling time} $t_*$, at which
$F_d-F(t_*) \sim \O(1) F_d$. For this class of systems,  it is expected that there is a parametric separation between these two time scales: $t_*\gg t_d$. 

\begin{figure}
\centering
\includegraphics[scale=0.55]{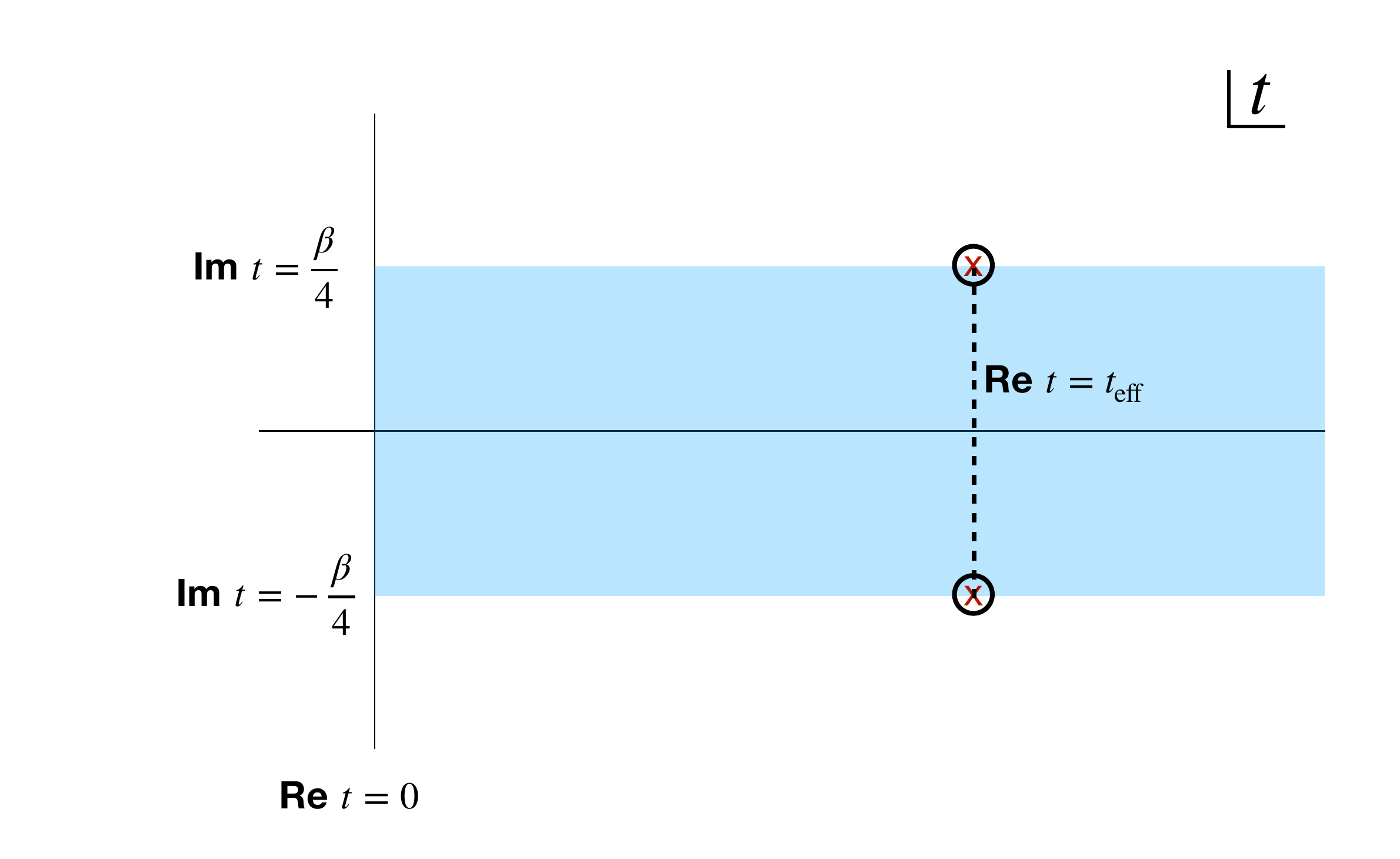}
\caption{ \label{figure:halfstrip} \small $F(t)$, as a function of complex $t$, is analytic in the shaded blue region. However, the extremal OTOC (\ref{eq:intro}) has singularities  at $t=t_{\rm eff}\pm i\beta/4$. These singularities can be removed by a simple $i \epsilon$-shift. }
\end{figure}

More generally, it is expected that this class of systems satisfies all the assumptions made in the derivation of the MSS chaos bound in \cite{Maldacena:2015waa} and the subleading chaos bounds in \cite{Kundu:2021qcx}. In particular,  the OTOC (\ref{eq:otoc}) has the following properties \cite{Maldacena:2015waa,Kundu:2021qcx}:
\begin{itemize}
\item[(i)] {\bf Analyticity}: $F(t)$ is an analytic function of $t$ in the half strip (see figure \ref{figure:halfstrip}): 
\be\label{halfstrip}
\mbox{Re}\ t >0 \qquad \text{and} \qquad -\frac{\beta}{4}\le \mbox{Im}\ t \le \frac{\beta}{4}\ .
\ee
\item[(ii)] {\bf Schwarz Reflection}: $F(t)$ obeys the Schwarz reflection condition 
\be\label{eq:SR}
\(F(t)\)^*=F\(t^*\)\ .
\ee
\item[(iii)] {\bf Boundedness}: $F(t)$ is bounded on the boundary  of the half strip (\ref{halfstrip})
\be\label{eq:positive}
|F(t\pm i \beta/4)|\le F_d \qquad  \text{for} \qquad  t\ge t_0 \ .
\ee  
\end{itemize}
The first two properties follow directly from the structure of the correlator (\ref{eq:otoc}) and hence they hold in general. On the other hand, property (iii) is more subtle.  The time scale $t_0$ in equation (\ref{eq:positive}) is the {\it factorization time}, which is defined as follows. It is the minimum time above which the time ordered thermal correlator $\tr \[y^2 W(t) V(0) y^2 V(0)W(t)\]$ approximately factorizes 
\be\label{eq:factorize}
\tr \[y^2 W(t) V(0) y^2 V(0)W(t)\]\approx F_d \qquad \text{for}\qquad t\ge t_0\ .
\ee
For the class of systems we are considering, the factorization time $t_0$ is expected to be larger than the dissipation time but much smaller than the scrambling time: $t_d<t_0\ll t_*$. Of course, $t_0$  can depend on the choice of operators in a specific quantum system. Nevertheless, for any simple Hermitian local operators $V$ and $W$, two time scales $t_0$ and $t_d$, both practically and conceptually, are not very different \cite{Kundu:2021qcx}.

It is very likely that properties (i)-(iii) are general enough to hold for even a larger class of chaotic systems.  For all these  quantum systems,  Maldacena, Shenker, and Stanford proved a universal bound in \cite{Maldacena:2015waa} on the rate of growth of $F(t)$:\footnote{To be precise, in \cite{Maldacena:2015waa} a stronger version of the boundedness condition (\ref{eq:positive})
\be\label{bounded:MSS}
|F(t)|\le F_d \qquad  \text{for} \qquad  \mbox{Re}\ t\ge t_0 \qquad  \text{and} \qquad  |\mbox{Im}\ t| \le \beta/4
\ee
was assumed to derive the MSS bound (\ref{eq:MSS}). This stronger assumption additionally requires that $|F(t)|\le F_d$ for $\mbox{Re}\ t=t_0$ and $| \mbox{Im}\ t|\le \beta/4$. However, this stronger boundedness condition is not actually necessary for the derivation of the MSS bound and its generalizations, as shown in \cite{Kundu:2021qcx}.}
\be\label{eq:MSS}
\frac{d}{dt}\(F_d-F(t)\) \le \frac{2\pi}{\beta}\(F_d-F(t)\)  \qquad \text{for}\qquad t\gg t_0\ .
\ee
For systems with a large separation between the factorization time and the scrambling time,  a clear signature of chaos is that the OTOC exhibits Lyapunov behavior
\be\label{Lyapunov0}
F_d-F(t)=\frac{c_1}{\N} e^{\lambda_L t}+\cdots \ , \qquad \text{for}\qquad t_*\gg t\gg t_0, t_d
\ee
where $c_1$ is a positive order one coefficient and $\N\gg 1$ is an effective measure of the number of degrees of freedom per site.  Hence, the bound (\ref{eq:MSS}) translates into  a  bound on the Lyapunov exponent $\lambda_L$ \cite{Maldacena:2015waa}
\be\label{MSS:Lyapunov}
\lambda_L \le \frac{2\pi}{\beta}\ .
\ee
This surprising bound on chaos, however, is not a mathematical coincidence, but part of a more general set of constraints that are contained in properties (i)-(iii). We review these general bounds below.

\subsection{Bounds}
The MSS bound (\ref{eq:MSS}) does not fully utilize properties (i)-(iii). For example, there are other bounds on subleading growing terms that are present in the OTOC. All these constraints can be organized systematically  by defining local {\it moments of the OTOC} \cite{Kundu:2021qcx}
\be\label{def:moments}
\mu_J\(t\)=e^{\frac{4\pi J}{\beta}t} \int_{t-i \frac{\beta}{4}}^{t+i \frac{\beta}{4}} dt' e^{-\frac{2\pi  }{\beta}\(t'-i \frac{\beta}{4}\)(2J+1)}  \(F(t')-F_d\)
\ee 
for real $t\ge t_0$ and any $J$. Note that the Schwarz reflection condition (\ref{eq:SR}) implies that moments $\mu_J(t)$ are real for integer $J$.

It was shown in \cite{Kundu:2021qcx} that in interacting unitary quantum systems with a large number of degrees of freedom and a Hamiltonian which is made out of finite products of simple operators, these moments must obey positivity, monotonicity, and log-convexity conditions:
\begin{align}
&\mu_J(t)>0\ ,\qquad \mu_{J+1}<\mu_J(t)\ ,\label{bound1}\\
&\mu_{J+1}(t)^2\le \mu_J\(t\)\mu_{J+2}\(t\)\label{bound2}\ ,
\end{align}
for $t\ge t_0$ and all integer $J\ge 0$.\footnote{Similar to the MSS bound, these bounds are actually valid, in general, for all OTOCs satisfying (i) analyticity, (ii) Schwarz reflection, and (iii) boundedness properties with a well-defined factorization time $t_0$. However, strictly speaking, the factorization condition (\ref{eq:factorize}) can break down for large $t\gg t_*$ because of Poincare recurrences. Hence, chaos bounds (\ref{bound1})-(\ref{bound1b}) are valid only up to some time scale which is smaller than the Poincare recurrence time of the system. This is true even for the MSS bound (\ref{eq:MSS}) \cite{Maldacena:2015waa}.}  Moreover, the moments are also bounded 
\be\label{bound1b}
\mu_J(t)<\frac{2\beta F_d}{\pi(2J+1)}e^{-\frac{2\pi }{\beta}t}
\ee
for real $t\ge t_0$ and integer $J\ge 0$. The MSS bound is contained in these more general constraints. In particular, the MSS bound is the leading constraint that one obtains from (\ref{bound1}). In addition, the above constraints also lead to bounds on subleading growths, as shown in \cite{Kundu:2021qcx}.

We wish to emphasize that conditions (\ref{bound1})-(\ref{bound1b}) lead to two types of constraints on $F(t)$: (I) inequalities that can be saturated and (II) strict inequalities. This observation will play an important role when we derive the extremally chaotic OTOC (\ref{eq:intro}).

\subsection{Maximal Chaos vs Extremal Chaos}
A quantum system is said to be maximally chaotic when its OTOCs exhibit a period of Lyapunov growth (\ref{Lyapunov0}) where the Lyapunov exponent saturates the MSS bound (\ref{MSS:Lyapunov})
\be\label{maximal}
F_d-F(t)=\frac{c_1}{\N} e^{\frac{2\pi}{\beta} t}+\cdots \ , \qquad \qquad \text{for}\qquad\qquad  t\gg t_0, t_d\ .
\ee
Theories of quantum gravity and their holographic duals are known to be maximally chaotic at the leading order in $1/\N$ \cite{Roberts:2014isa,Shenker:2013pqa,Shenker:2013yza,kitaev2014hidden,Shenker:2014cwa,Maldacena:2015waa}.

The OTOC must deviate significantly from (\ref{maximal}) for times comparable to the scrambling time $t_*=\frac{\beta}{2\pi} \ln \N$, since $|F(t)|$ is bounded from above. To see that first consider some arbitrary reference time $t_*\gg t_{\rm ref}\gg t_0$. From (\ref{maximal}) we find that $|F(t_{\rm ref}+i \tau)| \le F_d +\frac{\O(1)}{\N^2}$, where $|\tau|< \beta/4$. On the other hand, the boundedness condition (\ref{eq:positive}) implies that $|F(t)|$ is also bounded on the boundary $|\mbox{Im}\ t|=\beta/4$ for $\mbox{Re}\ t \ge t_{\rm ref}$. Moreover, it was shown in \cite{Maldacena:2015waa} that $|F(t)|$ is bounded by some finite constant everywhere in the interior. Hence, the Phragmén-Lindelöf principle implies that any analytic completion of (\ref{maximal}) must obey
\be\label{eq:bounded}
|F(t)| \le F_d +\frac{\O(1)}{\N^2}
\ee
everywhere in the interior of the half strip for $\mbox{Re}\ t\gg t_0$.\footnote{In contrast, the condition (\ref{eq:bounded}) was assumed to be exact in \cite{Maldacena:2015waa}. This clearly reflects the fact that our boundedness condition (\ref{eq:positive}) is weaker than the boundedness assumption (\ref{bounded:MSS}) of \cite{Maldacena:2015waa}. }

In fact, there has to be corrections (however small) to the maximally chaotic OTOC (\ref{maximal}) even for $t\ll t_*$. In order to see that, we can compute the moments (\ref{def:moments}) associated with the OTOC (\ref{maximal}):
\begin{align}\label{qg:moments}
\mu_0(t)=\frac{c_1 \beta}{2\N}>0\ , \qquad \mu_{J\ge 1}(t)=0
\end{align}
for integer $J$. This is at odds with (\ref{bound1}) implying that the term $e^{\frac{2\pi}{\beta} t}$ alone, in any time duration, is inconsistent with properties (i)-(iii) of the OTOC. So, there has to be correction terms.  The maximally chaotic OTOC (\ref{maximal}) can be analytically completed in many different ways. We introduce the extremally chaotic OTOC as a very specific analytic completion of  the maximally chaotic OTOC (\ref{maximal}), saturating all the chaos bounds obtained from (\ref{bound1}) and (\ref{bound2}) that can be saturated.

As mentioned in the introduction, the extremal OTOC $F_{\rm ext}(t)$ is defined as the OTOC with the following properties: (1) it saturates all the chaos bounds obtained from (\ref{bound1}) and (\ref{bound2}) that allow saturation, (2) it coincides with (\ref{maximal}) in some duration within $t_0< t< t_*$, (3) it is an analytic function inside the entire half-strip $\{t\in \mathbb{C}|\ \mbox{Re}\ t\ge t_0\ \text{and}\ |\mbox{Im}\ t|< \frac{\beta}{4} \}$, obeying the Schwarz reflection property.

The constraints (\ref{bound1}) and (\ref{bound2}) lead to an infinite set of chaos bounds that allow saturation. To begin with, it is not obvious whether these bounds can all be saturated by any non-trivial OTOC in a consistent way.  So, it is indeed satisfying to find that the OTOC (\ref{eq:intro}) saturates all these chaos bounds, both leading and subleading. This OTOC is not only bounded for real $t$  even for times large compared to the scrambling time but also automatically satisfies  conditions (3). Furthermore, the constraints (\ref{bound1}) and (\ref{bound2}) guarantee that $F_{\rm ext}(t)$ is unique up to terms that decay exponentially for $t\gg t_0$.

\section{Extremal OTOC}\label{sec:extremal}
In this section, we derive the extremal OTOC (\ref{eq:intro}) and argue for its uniqueness. First, we give a simple derivation by considering OTOCs that can be written as a sum of Lyapunov growths: $F(t)\sim \sum_i e^{\lambda_i t}$. Later, we will provide a more rigorous derivation.

\subsection{A Lyapunov Expansion of the OTOC}
As we have explained in the last section, the maximally chaotic OTOC (\ref{maximal}) must contain additional correction terms which we parametrize as follows:
\be\label{eq:para}
F_d - F(t)= \frac{1}{\N}\(c_1e^{\frac{2\pi}{\beta} t}+ c_2  e^{\lambda_2 (t-t_f)}+\cdots\) \qquad t_0\ll \mbox{Re}\ t\le t_f\ ,
\ee
where $t_f$ is a new time scale the physical meaning of which will be clear later. The constraints (\ref{bound1}) now impose \cite{Kundu:2021qcx} 
\be\label{MSS2}
\lambda_2 \le \frac{6\pi}{\beta}\ .
\ee
At first sight, it might be surprising that the above bound can be saturated since it follows from a set of strict inequalities (\ref{bound1}). So, the saturation of the bound $\lambda_2 = \frac{6\pi}{\beta}$ requires some discussion. In this case, one can check that $\mu_{J\ge 2}(t)=0$ for integer $J$. This is  inconsistent with  constraints  (\ref{bound1}), however, that does not mean $\lambda_2 = \frac{6\pi}{\beta}$ is not allowed. It only means that the saturation of (\ref{MSS2}) necessarily requires additional correction terms in (\ref{eq:para}) that generate positive contributions for $\mu_{J\ge 2}(t)$. 

So, we impose that the OTOC (\ref{eq:para}) saturates the bound (\ref{MSS2}) as well, fixing $\lambda_2 = \frac{6\pi}{\beta}$. Similarly, we can  repeat the preceding argument again and again by adding exponential correction terms. At each step, the upper bound on the Lyapunov exponent increases by $\frac{4\pi}{\beta}$ \cite{Kundu:2021qcx}. Hence, we write the extremal OTOC as an infinite series
\be\label{eq:specialcase}
F_d - F(t)= \frac{1}{\N}e^{\frac{2\pi}{\beta} t}\sum_{n=0}^{\infty}c_{n+1} e^{\frac{4n\pi}{\beta} (t-t_f)}\ , \qquad t_0\ll \mbox{Re}\ t\le t_f\ .
\ee 
From the MSS bound, we expect that $|c_2|,|c_3|,\cdots<c_1$ because the leading growing term should not violate (\ref{MSS:Lyapunov}). However, to begin with we will not assume any such restrictions on the $c$-coefficients. Rather, they will automatically follow from constraints (\ref{bound1}) and (\ref{bound2}). This should not be surprising since the MSS bound is contained in (\ref{bound1}).

Let us emphasize that even though the OTOC (\ref{eq:specialcase}) saturates an infinite number of bounds, there are additional (infinitely many) constraints from conditions (\ref{bound1}) and (\ref{bound2}). Importantly, some of the remaining constraints also allow saturation. It was shown in \cite{Kundu:2021qcx} that the OTOC (\ref{eq:specialcase}) is consistent with bounds (\ref{bound1}) and (\ref{bound2}) for $ \mbox{Re}\ t\le t_f$ if and only if 
\begin{align}
&(-1)^{n-1} c_{n}>0\ , \qquad |c_{n+1}|<|c_n|\ ,\label{bound3}\\
&c_{n+1}^2\le c_n c_{n+2}\ ,\label{bound4}
\end{align}
for all integer $n\ge 1$.\footnote{These constraints are closely related to analogous positivity, monotonicity, and log-convexity conditions of  certain CFT Regge correlators  \cite{Kundu:2020gkz,Kundu:2021qpi}. These CFT conditions follow directly from basic properties of Lorentzian correlators and can be regarded as causality constraints.} We obtain the extremally chaotic OTOC by saturating bounds (\ref{bound4}) without violating any of the strict inequalities (\ref{bound3}), yielding  
\be
F_{\rm ext}(t)= F_d - \frac{c_1 e^{\frac{2\pi}{\beta} t}}{\N}\sum_{n=0}^\infty (-1)^n \varepsilon^n  e^{\frac{4n \pi}{\beta} (t-t_f)}\ ,
\ee
where, $\varepsilon\equiv |c_2|/c_1<1$. The above sum converges only for $\varepsilon e^{\frac{4 \pi}{\beta} (t-t_f)}<1$, however, it can be analytically continued beyond the regime of convergence. In particular, after defining a new time scale $t_{\rm eff}= t_f-\frac{\beta}{4\pi} \ln \varepsilon>t_f$, the above expansion can be ressumed,  obtaining
\be\label{fmax1}
F_{\rm ext}(t)= F_d -   c_1  \frac{e^{\frac{2\pi}{\beta}( t-t_*)}}{1+ e^{\frac{4 \pi}{\beta} (t-t_{\rm eff})}}\equiv F_d -\frac{c_1}{\N} \mathcal{F}_{\rm ext}(t;t_{\rm eff})\ .
\ee
The OTOC $F_{\rm ext}(t)$ is well-defined even when $t>t_{\rm eff},t_*$. 

So, the function $\mathcal{F}_{\rm ext}(t;t_{\rm eff})$ is completely fixed only up to a theory dependent time scale $t_{\rm eff}$. Physically, $t_{\rm eff}$ represents the time of maximum scrambling of the initial perturbation since the extremal OTOC $F_{\rm ext}(t)$ has a global minimum at $t=t_{\rm eff}$. However, the time scale $t_{\rm eff}$, in general, is independent of the scrambling time $t_*$. Furthermore,  $t_{\rm eff}$ is not required to be parametrically smaller than the scrambling time. Nevertheless, $t_{\rm eff}$ is not completely free of constraints. The boundedness condition (\ref{eq:bounded}) on the real line does impose an upper bound:
\be
t_{\rm eff}\le t_* +\frac{\beta}{2\pi}\ln \(\frac{4F_d}{c_1}\)+\beta \frac{\O(1)}{\N^2}\ .
\ee
Clearly, the OTOC $F_{\rm ext}(t)$ is an analytic function obeying the Schwarz reflection condition (\ref{eq:SR}) inside the half-strip $\{t\in \mathbb{C}|\ \mbox{Re}\ t\ge t_0\ \text{and}\ |\mbox{Im}\ t|< \frac{\beta}{4} \}$. On the other hand, from our derivation, it is unclear whether $F_{\rm ext}(t)$ is unique. After all, this derivation relies heavily on our initial assumption that the extremally chaotic OTOC can be written as $F(t)\sim \sum_i e^{\lambda_i t}$.  We will close this loophole next by providing a more rigorous argument.

\subsection{Uniqueness of the Extremal OTOC}
We now show that the extremal OTOC (\ref{eq:intro}) is unique up to terms that decay for $t\gg t_0$ by using tools developed in \cite{Kundu:2021qcx}. We start with the dispersion relation of \cite{Kundu:2021qcx} 
 \begin{align}\label{eq:late}
F_d-F(t)= \frac{2}{\beta} e^{-\frac{2\pi (t-t_0)}{\beta}}& \int^{\frac{\beta}{4}}_{-\frac{\beta}{4}} d\tau \frac{e^{\frac{2\pi i \tau}{\beta}}\(F_d-F(t_0+i \tau)\)}{1- e^{\frac{4\pi i \tau}{\beta}}e^{-\frac{4\pi (t-t_0)}{\beta}}}- \frac{2}{\beta} e^{\frac{2\pi}{\beta}t}\int_{t_0}^\infty dt' \frac{\mu_0'(t')}{1+e^{\frac{4\pi }{\beta}(t-t')}}\ ,
\end{align}
written in terms of the primary moment $\mu_0(t)$. Any OTOC that satisfies properties (i)-(iii) of section \ref{sec:chaos} can be written in this form for $\mbox{Re}\ t> t_0$ and $|\mbox{Im}\ t|< \beta/4$. One advantage of writing the OTOC in this form is that the first term decays as $e^{-\frac{2\pi (t-t_0)}{\beta}}$ for $\mbox{Re}\ t\gg t_0$. So, all we need to do is to determine $\mu_0(t)$ associated with the extremal chaotic OTOC. 

At this point, it is useful to saturate the log-convexity bound (\ref{bound2}) first. The most general $\mu_J(t)$ that saturates the log-convexity bound (\ref{bound2})  is
\be\label{eq:saturate}
\mu_J(t)= \mu_0(t) e^{-\frac{4\pi J}{\beta} g(t)}
\ee
for integer $J\ge 0$, where $g(t)$ is a real function independent of $J$. This automatically satisfies bounds (\ref{bound1}), provided both $\mu_0(t)$ and $g(t)$ are positive functions. Set of functions (\ref{eq:saturate}), for arbitrary $g(t)$, does not represent a well-defined set of moments.  For example, the moments (\ref{def:moments}) must satisfy the consistency condition \cite{Kundu:2021qcx}
\be\label{eq:consistency}
\mu_J'(t)-\frac{4\pi J}{\beta} \mu_J(t)=\mu_0'(t)
\ee
for all positive integer $J$. This consistency condition is actually highly constraining. This becomes obvious when we apply it to (\ref{eq:saturate}):
\be\label{eq:consistency2}
\mu_0'(t)e^{-\frac{4\pi J}{\beta} g(t)}- \frac{4\pi J}{\beta}e^{-\frac{4\pi J}{\beta}g(t) }\(g'(t)+1\)\mu_0(t)=\mu_0'(t)\ .
\ee 
The left-hand-side must be independent of $J$ implying that $g'(t)=-1$ when $\mu_0(t)$ is non-zero. So, we find 
\be
g(t)=t_{\rm eff}-t\ ,
\ee
where, $t_{\rm eff}$ is a constant. This constant should be regarded as the effective scrambling time (or a gap time scale), as we discuss below. Note that the consistency condition (\ref{eq:consistency2}) is still not satisfied, unless
\be\label{sol:mu0}
\mu_0'(t)=0 \qquad \text{when}\qquad t\neq t_{\rm eff} \ .
\ee
However, $\mu_0(t)$ cannot be the same everywhere. From the definition (\ref{def:moments}), we know that $ \mu_J(t)$ goes to zero as $t\rightarrow \infty$, for all integer $J\ge 0$. On the other hand, even a brief period of maximal chaos requires $\mu_0(t)$ to be non-zero in that interval. Hence, for extremal chaos, as defined in this paper, $\mu_0(t)$ must be a piecewise constant function of time. Alternatively, the relation (\ref{eq:consistency}) implies that 
\be
\mu_J( t)=- e^{\frac{4\pi J}{\beta}t}\int_{t}^\infty dt' e^{-\frac{4\pi J}{\beta}t'} \mu_0'(t')\ ,
\ee
where $\mu_0'(t)$ satisfies (\ref{sol:mu0}). The above integral can be non-zero for $J\ge 1$ only if $\mu_0'(t)\propto \delta(t-t_{\rm eff})$. Therefore, there is a unique solution that is consistent with (\ref{eq:saturate}) and (\ref{eq:consistency}):
\be\label{eq:extchaos}
\mu_J(t)=\mu e^{\frac{4\pi J}{\beta}(t-t_{\rm eff})} \Theta\(t_{\rm eff}-t\)\ ,
\ee
where $\mu$ is a positive constant. Moreover, a period of maximal chaos requires $t_{\rm eff}\gg t_0$, implying $t_{\rm eff}$ can also be interpreted as a gap time scale. 

We now utilize the representation (\ref{eq:late}) to obtain the unique extremally chaotic OTOC for $t\gg t_0$:
\be\label{otoc:max}
F_{\rm ext}(t)= F_d -   \frac{2\mu}{\beta}  \frac{e^{\frac{2\pi}{\beta}t}}{1+ e^{\frac{4 \pi}{\beta} (t-t_{\rm eff})}}+\O\(e^{-\frac{2\pi (t-t_0)}{\beta}}\)
\ee
where we identify $\mu= \frac{\beta c_1}{2\N}$. This agrees with our previous derivation, upto terms that decay at late times.  

The extremal OTOC $F_{\rm ext}(t)$ decays fast for $t_0\ll t\ll t_{\rm eff}$, saturating the MSS bound. It reaches its minimum value at $t=t_{\rm eff}$. So, in some sense $t_{\rm eff}$ is an effective scrambling time, even though it is independent of the traditional scrambling time  $t_*=\frac{\beta}{2\pi}\ln \N$. Importantly, $F_{\rm ext}(t)$ is well-defined even for $t>t_{\rm eff},t_*$, provided $|\mbox{Im}\ t|< \beta/4$. Above $t>t_{\rm eff}$, the OTOC $F_{\rm ext}(t)$ increases monotonically, indicating that information of the initial perturbation is not completely lost. The information can be fully recovered in the limit $t\rightarrow \infty$ at which $F_{\rm ext}(t)\rightarrow F_d$. 
 
Let us now compare the extremal OTOC (\ref{otoc:max}) with thermal OTOCs of large $N$ holographic CFTs. For example, in 2d CFTs with a large central charge $c$ (and a sparse spectrum), $F(t)$ can be computed beyond the leading $\frac{1}{c}$ term for certain heavy operators \cite{Roberts:2014ifa,Shenker:2014cwa}. These OTOCs asymptote to zero for $t>t_*$, indicating a loss of information. Similarly, $F(t)$ in the Schwarzian theory also asymptotes to zero for $t> t_*$ \cite{Maldacena:2016upp,Lam:2018pvp}.

There is an important caveat that we must address. The extremal solution (\ref{eq:extchaos}) appears to be inconsistent with the bound (\ref{bound1}) for $t> t_{\rm eff}$. This inconsistency stems from the fact that $F_{\rm ext}(t)$ has singularities on the boundary of the half strip (\ref{halfstrip}). We will argue that this non-analyticity simply means $F_{\rm ext}(t)$ should be interpreted as a distribution.  
 
 \subsection{Non-Analyticity}

We notice that the OTOC (\ref{otoc:max}) has simple poles at $t=t_{\rm eff}\pm i\beta/4$. Strictly speaking, the expression (\ref{otoc:max}) is not valid on the boundary $|\mbox{Im}\ t|= \beta/4$, since it has been derived from the dispersion relation (\ref{eq:late}). Nevertheless, the value of the OTOC on the boundary of the half-strip (\ref{halfstrip}) can also be determined directly from the extremal solution (\ref{eq:extchaos}). In particular, from \cite{Kundu:2021qcx} (see appendix B) we find that for real $t$:
  \begin{align}
& \mbox{Re}\ F_{\rm ext}\(t+i\frac{\beta}{4}\)=F_d-\frac{\beta c_1}{4\N}e^{\frac{2\pi}{\beta}t_{\rm eff}}\delta(t-t_{\rm eff})\ ,\label{boundary1}\\
&\mbox{Im}\ F_{\rm ext}\(t+i\frac{\beta}{4}\)= -\frac{c_1}{\N}  \frac{e^{\frac{2\pi}{\beta}t}}{1-e^{\frac{4\pi }{\beta}(t-t_{\rm eff})}}+\O\(e^{-\frac{2\pi (t-t_0)}{\beta}}\)\ ,\label{boundary2}
 \end{align}
 implying that the extremal OTOC is indeed singular at $t=t_{\rm eff}\pm i\beta/4$.
 We are familiar with these kinds of singularities in QFT. Mathematically, these singularities can be easily removed by a standard $i \epsilon$-shift\footnote{Consequently, the extremal OTOC (\ref{otoc:max}) is not consistent with the boundedness condition (\ref{eq:bounded}) near the boundary of the half-strip (\ref{halfstrip}). This can also be resolved by the $i\epsilon$-shift, leading to a lower bound on $\epsilon$.} 
 \be\label{reg:OTOC}
 F_{\rm ext}^\epsilon\(t\pm i\frac{\beta}{4}\)= F_d \mp  i \frac{c_1}{\N}  \frac{e^{\frac{2\pi}{\beta}t}}{1- e^{\frac{4 \pi}{\beta} (t-t_{\rm eff})\mp i \epsilon}}+\O\(e^{-\frac{2\pi (t-t_0)}{\beta}}\)\ ,
 \ee
 where $t\gg t_0$ is real and $\epsilon>0$ is small. We recover boundary values of the extremal OTOC (\ref{boundary1}) and (\ref{boundary2}) from $ F_{\rm ext}^\epsilon$ by taking the limit $\epsilon\rightarrow 0$. 
 
Equivalently, the role of the $i\epsilon$-shift can be stated in the following way.  A well-behaved OTOC obeying properties (i)-(iii) of section \ref{sec:chaos} cannot have $\mu_0'(t)\propto \delta(t-t_{\rm eff})$ since $|F(t)|$ is bounded in the half-strip $\S=\{t\in \mathbb{C}|\ \mbox{Re}\ t\ge t_0\ \text{and}\ |\mbox{Im}\ t|\le \frac{\beta}{4} \}$.  The $i\epsilon$-shift in (\ref{reg:OTOC}) replaces this delta function by a smooth but a narrow function. In fact, in physical systems there is a lower bound on $\epsilon$ which may not always be infinitesimally small.  We will discuss this in section \ref{sec:physical}.

Of course, the  $i\epsilon$-shift (\ref{reg:OTOC}) is a very specific small deformation of the extremal OTOC  $F_{\rm ext}(t)$, making it regular at $t=t_{\rm eff}\pm i\beta/4$. It is actually possible to study deformations, large or small, of the extremal OTOC in a very general way. This can be achieved by treating the extremal OTOC as a distribution, as we discuss next. 

 \section{General Deformations of Extremal Chaos}\label{sec:KL}
 \subsection{Spectral  Representation of OTOC}
\begin{figure}
\centering
\includegraphics[scale=0.55]{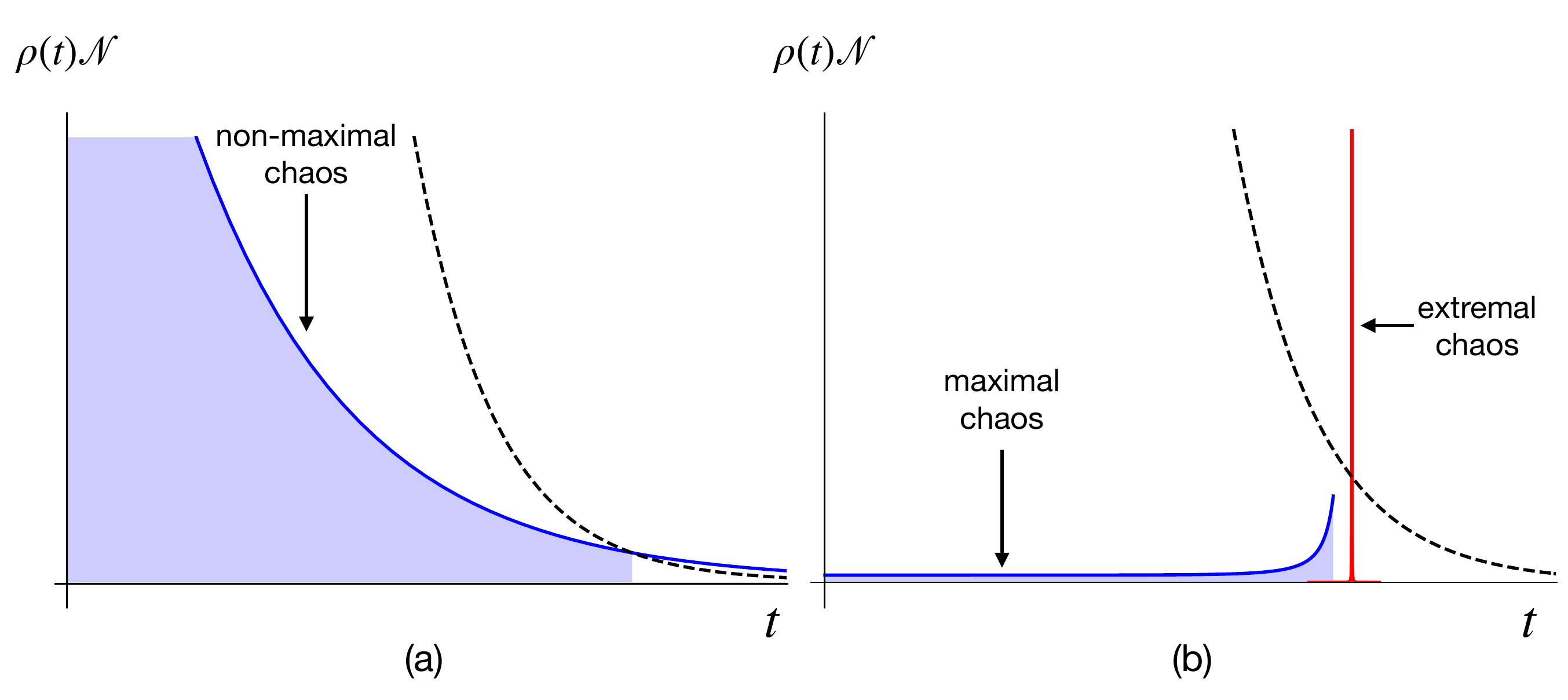}
\caption{ \label{figure:density} \small Density functions (\ref{eq:inversion}) associated with various chaotic systems are shown here schematically. The dashed black line represents the bound (\ref{rho:bound}). Figure (a): The blue line is the density function for a period of Lyapunov growth with $0<\lambda_L< \frac{2\pi}{\beta}$. This density function becomes inconsistent with the bound (\ref{rho:bound}) for large $t$, where correction terms must become significantly large so that the density function decays faster.  Figure (b): The blue line represents a typical density function associated with a period of maximal chaos. Over the same duration of time the density function $\rho(t)\approx 0$, however, it starts to grow around some time scale $t=t_{\rm eff}$. The red line represents the delta-function density associated with the extremally chaotic OTOC. }
\end{figure}

We begin this section with an observation that the extremally chaotic OTOC gives a K\"{a}llen-Lehmann-type representation of any general OTOC (not necessarily maximally chaotic in any duration). In particular, any OTOC obeying properties (i)-(iii) of section \ref{sec:chaos} can be written as 
 \be\label{eq:KL}
 F_d-F(t)=\int_{t_0}^\infty dt' \mathcal{F}_{\rm ext}(t;t')\rho(t')+\O\(e^{-\frac{2\pi (t-t_0)}{\beta}}\)
 \ee
for $\mbox{Re}\ t\gg t_0$ and $|\mbox{Im}\ t|< \beta/4$, where $\mathcal{F}_{\rm ext}(t;t')$ is the extremal function as defined in (\ref{eq:intro}). The function $\rho(t)$ parallels the spectral density of the original K\"{a}llen-Lehmann representation, and hence it can be thought of as a  density function of chaos. The spectral representation (\ref{eq:KL}) follows directly from the dispersion relation (\ref{eq:late}) once we notice that the kernel in the second term of (\ref{eq:late}) is precisely $\mathcal{F}_{\rm ext}(t;t')$. Furthermore, the dispersion relation (\ref{eq:late}) provides a simple inversion formula for the density function 
\be\label{eq:inversion}
\rho(t)=-\frac{2}{\beta}\mu_0'(t)=\frac{4}{\beta}e^{-\frac{2\pi t}{\beta}}\(F_d-\mbox{Re}\ F\(t+i\frac{\beta}{4}\)\)\ .
\ee
This relation implies that the density function must have the following properties for $t\ge t_0$:
\begin{itemize}
\item{It is real and non-negative: $\rho(t)\ge 0$.}
\item{It is smooth (infinitely differentiable).}
\item{It is bounded: 
\be\label{rho:bound}
\rho(t)\le \frac{8}{\beta}e^{-\frac{2\pi t}{\beta}}F_d
\ee
and hence $\rho(t\rightarrow \infty)\rightarrow 0$. 
}
\item{A period of maximal chaos over which the MSS bound is saturated necessarily requires $\rho(t)\approx 0$ over the same time duration.\footnote{Note that $\rho(t)$ can only have isolated zeroes since $F(t)$ is analytic in the domain (\ref{halfstrip}). On the other hand, if the OTOC saturates the MSS bound exactly, then $\rho(t)=0$. So, this again implies that a term $e^{\frac{2\pi t}{\beta}}$ in the OTOC always comes with correction terms (see figure \ref{figure:density}).} }
\end{itemize}
So, any OTOC can be thought of as a deformation of the extremal OTOC (\ref{eq:intro}) in a very precise way. Moreover, the representation (\ref{eq:KL}) has the important advantage that all the chaos bounds, leading or subleading, are automatically satisfied, provided $\rho(t)$ obeys the above conditions. 

Interestingly, the representation (\ref{eq:KL}) can be mapped to a statistical physics problem of a {\it Fermi gas}. In particular, $1-\exp(-\frac{2\pi t}{\beta})\mathcal{F}_{\rm ext}(t;t')\equiv f_{FD}(t;t')$ is exactly the {\it Fermi-Dirac distribution} once we substitute $t'\rightarrow E$, $t\rightarrow \mu$, and $\beta \rightarrow 2\pi k_B T$. Hence, the representation (\ref{eq:KL}) can be viewed as an integrals of the  Fermi-Dirac distribution computing the number density in a Fermi gas. Conceptually, this is consistent with our interpretation of $\rho(t)$ as a density. More practically, this identification enables us to evaluate (\ref{eq:KL}), exactly or approximately, by utilizing various mathematical tools available for integrals of the  Fermi-Dirac distribution.

The density function $\rho(t)$ is obtained from a given OTOC-behavior. Typically, OTOCs exhibit various classes of functional behaviors: exponential (chaotic), power-law (not quite chaotic), constant (integrable). All these statements translate straightforwardly to some specific features of $\rho(t)$. For example, consider a period of exponential growth
\be\label{Lyapunov}
F(t)=F_d-\frac{c_1}{\N} e^{\lambda_1 t}+\cdots \ ,  
\ee
for $t\gg t_0$, where $c_1>0$ and $\lambda_1< \frac{2 \pi  }{\beta }$. As explained above, this OTOC can be written as (\ref{eq:KL}) where the density function is given by
\be
\rho(t)=\frac{4 c_1}{\N \beta} e^{t \left(-\frac{2 \pi  }{\beta }+\lambda_1 \right)}\cos \left(\frac{\beta  \lambda_1 }{4}\right)
\ee
which is real and positive. The density function decays with time, however, it does not decay fast enough when $\lambda_1>0$, as shown in figure \ref{figure:density}. In particular, this density function becomes inconsistent with the bound (\ref{rho:bound}) near the scrambling time, implying a breakdown of the approximation (\ref{Lyapunov}).

\subsection{Extremal Chaos in Physical Systems}\label{sec:physical}
\begin{figure}
\centering
\includegraphics[scale=0.55]{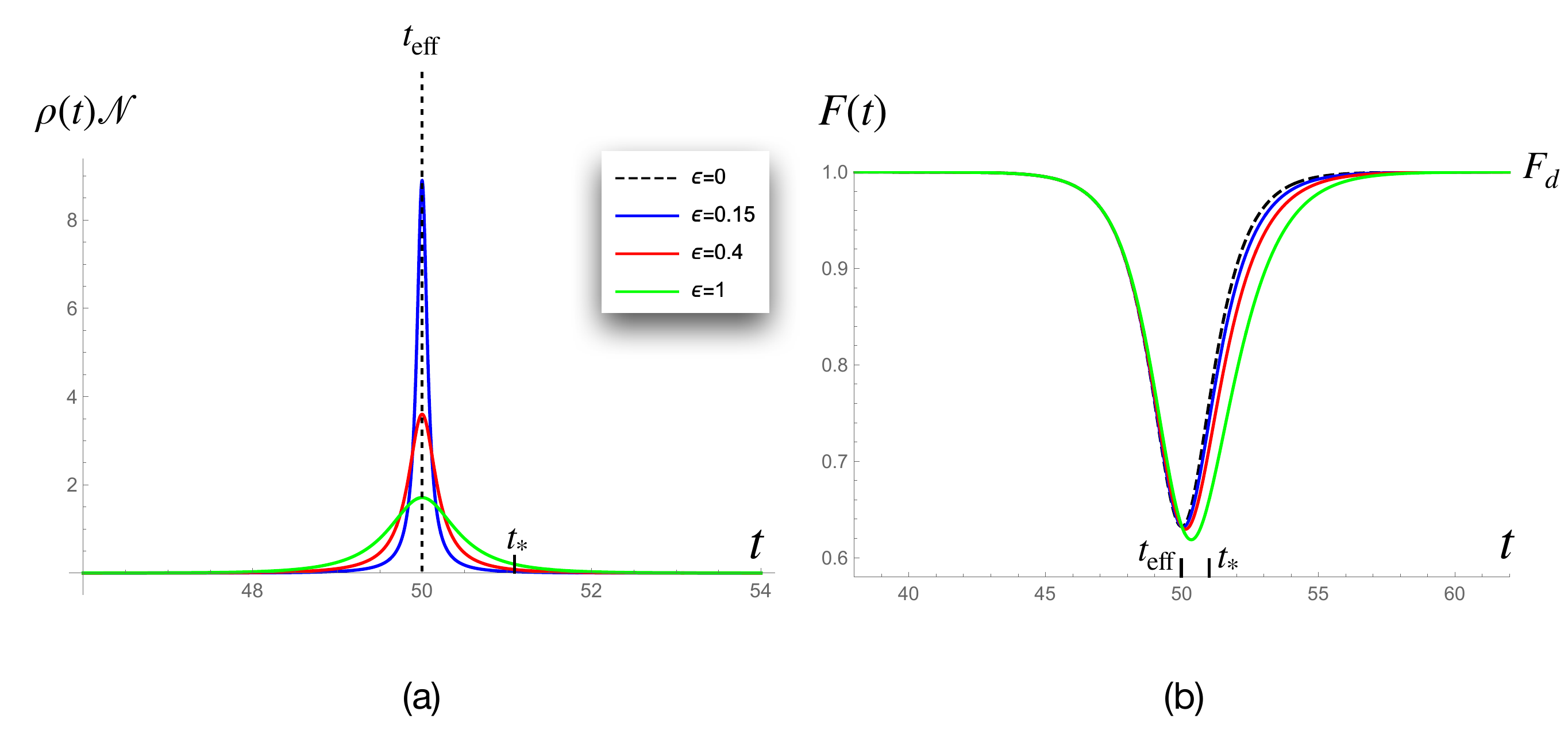}
\caption{ \label{figure:reg} \small The regularized density functions (a) and the associated OTOCs (b) are shown for various values of $\epsilon$. For the plot, we have chosen $\beta=2\pi, F_d=1, c_1=2, t_{\rm eff}=50$ and $t_*=51$. For this set of parameters, the bound (\ref{epsilon:bound}) requires that $\epsilon \ge 0.1415$. The dashed black lines represent the unregularized OTOC (\ref{eq:intro}). The regularized OTOC differs only slightly from the unregularized OTOC even when $\epsilon=1$.}
\end{figure}

In this language, extremal chaos is characterized by a ``single-particle state":  $\rho(t)=\frac{c_1}{\N}\delta(t-t_{\rm eff})$ (see figure \ref{figure:density}).  Clearly, this is in tension with the boundedness property (\ref{rho:bound}) of the density $\rho(t)$. This tension is a manifestation of the non-analyticity of the extremally chaotic OTOC on the boundary: $\mbox{Im}\ t=\pm \beta/4$. A natural resolution of this tension is to remove these singularities by performing the $i\epsilon$-regularization of (\ref{reg:OTOC}). By using (\ref{eq:inversion}) we find that this $i\epsilon$-regularization essentially replaces the delta function by a narrow distribution (see figure \ref{figure:reg}):
\be\label{rho:extremal}
\rho(t)=\frac{4\tilde{c}_1}{\N \beta} \frac{e^{\frac{4 \pi}{\beta} (t-t_{\rm eff})}\sin \epsilon}{\(1- e^{\frac{4 \pi}{\beta} (t-t_{\rm eff})+i \epsilon}\)\(1- e^{\frac{4 \pi}{\beta} (t-t_{\rm eff})-i \epsilon}\)}
\ee
up to terms that decay for $t\gg t_0$. Note that we are using a normalization $\tilde{c}_1\propto c_1$ such that the coefficient of the maximally chaotic term $e^{\frac{2\pi t}{\beta}}$ remains unchanged $c_1/\N$, even when $\epsilon$ is finite. So, the proportionality constant $\tilde{c}_1/ c_1$, which we will determine later, depends on $\epsilon$ and approaches 1 in the limit $\epsilon \rightarrow 0$.\footnote{Alternatively, one may normalize (\ref{rho:extremal}) by keeping the integral $\int_{t_0}^\infty dt \rho(t)$ fixed (and independent of $\epsilon$). These two normalizations, however, are equivalent since the coefficient of the term $e^{\frac{2\pi t}{\beta}}$ is exactly this integral. This is actually true in general, as we will discuss in section \ref{sec:maximal}.}  

The bound (\ref{rho:bound}) now imposes a lower bound on $\epsilon$: 
\be\label{epsilon:bound}
\tan \(\frac{\epsilon}{2}\) \ge \frac{\tilde{c}_1}{4F_d}e^{\frac{2\pi }{\beta}\(t_{\rm eff}-t_*\)}\ .
\ee
In general, $t_{\rm eff}$ is independent of the scrambling time $t_*$.  However, in a strongly chaotic system $F_d-F(t\sim t_{\rm eff})\sim \O(1) F_d$, implying these two time scales cannot be parametrically separated. 

So, physical systems can be extremally chaotic only approximately where the density function $\rho(t)$ is a narrow peak at $t=t_{\rm eff}\gg t_0$ with width $\Delta t_{\rm eff}\sim \epsilon \frac{\beta}{2\pi}$. The associated OTOC is now an analytic function everywhere in the half-strip (\ref{halfstrip})  that saturates the MSS bound in the regime $t_0\ll t\ll t_{\rm eff},t_*$. This represents a small deformation, even when $\epsilon$ is finite, since the moments $\mu_J(t)$ asymptote to the extremal solution (\ref{eq:extchaos}) away from the time scale $t_{\rm eff}$. As a consequence, the resulting OTOC  can be well-approximated by the extremal OTOC (\ref{eq:intro}) in the regime $|t-t_{\rm eff}|\gg \frac{\beta}{2\pi}$.

We can determine the exact OTOC even when $\epsilon$ is finite (and real). This is a straightforward exercise because of the spectral representation (\ref{eq:KL}). In particular, for $t,t_*, t_{\rm eff}\gg t_0$ we obtain 
\be\label{otoc:reg}
F_{\rm ext}^{\rm reg}(t)=F_d-\frac{\tilde{c}_1}{\N}e^{\frac{2\pi t}{\beta}} \mbox{Re}\(\frac{1-\frac{\epsilon}{\pi}-\frac{4 i }{\beta }(t-t_{\rm eff})}{1+ e^{\frac{4 \pi}{\beta} (t-t_{\rm eff})- i \epsilon}}\)
\ee
for real $t$. When $\mbox{Im}\ t \neq 0$, the OTOC can be obtained from the spectral representation (\ref{eq:KL}) exactly the same way. This OTOC is now an analytic function everywhere in the half-strip (\ref{halfstrip}) obeying the Schwarz reflection (\ref{eq:SR}) and the boundedness (\ref{eq:positive}) conditions. Note that the regularized OTOC (\ref{otoc:reg}) has the same qualitative features as the extremal OTOC (\ref{otoc:max}), differing only slightly even when $\epsilon$ is order 1 (see figure \ref{figure:reg}). 

We now focus on the regime $t_0\ll t\ll t_{\rm eff},t_*$, in which the regularized OTOC (\ref{otoc:reg}) is well-approximated by
\be\label{otoc:approx}
F_{\rm ext}^{\rm reg}(t)=F_d-\frac{\tilde{c}_1 \(1-\frac{\epsilon}{\pi}\)}{\N}e^{\frac{2\pi t}{\beta}}\(1-e^{\frac{4 \pi}{\beta} (t-t_{\rm eff})}\(\cos \epsilon -\frac{4(t-t_{\rm eff})\sin \epsilon}{\beta \(1-\frac{\epsilon}{\pi}\)}\)+\cdots\)\ .
\ee
From the above limit, we find  
\be
c_1=\tilde{c}_1 \(1-\frac{\epsilon}{\pi}\)
\ee
such that the coefficient of the term $e^{\frac{2\pi t}{\beta}}$ remains unchanged $c_1/\N$.

The OTOC (\ref{otoc:reg}) is an analytic completion of the leading Lyapunov behavior (\ref{maximal}). This analytic completion asymptotes to the extremally chaotic OTOC  (\ref{eq:intro}) away from the time scale $t_{\rm eff}$. In theories of quantum gravity and their holographic duals, the scrambling time is determined by the Newton constant $G_N$. On the other hand, the time scale $t_{\rm eff}$, when independent, can be thought of as the analog of the string scale and the OTOC (\ref{otoc:reg}) can be interpreted as a ``tree-level" analytic completion of maximal chaos.

 \subsection{Late-Time Corrections from Long Tails}
 The $i\epsilon$ regularization makes the extremal OTOC  analytic even on the boundary of the half-strip (\ref{halfstrip}) while preserving all the qualitative features on the real line. For example, the OTOC (\ref{otoc:reg}) asymptotes to $F=F_d$ in the limit $t\rightarrow \infty$ for all $\epsilon$. This fact depends heavily on the late-time behavior of the density function (\ref{rho:extremal})
 \be\label{densityfunction_late}
 \rho(t> t_{\rm eff}) \approx \frac{4\tilde{c}_1 \sin \epsilon}{\N \beta}  e^{\frac{4 \pi}{\beta} (t_{\rm eff}-t)}\ .
 \ee
This density function becomes very small for $t>t_*$ and hence other effects can dominate at very late times. In particular, parametrically   $ \rho(t)\ll 1/\N$ for $t-t_{\rm eff}\gg \beta$ and hence in this regime higher-order $1/\N$ corrections to (\ref{rho:extremal}) can be important. However, as we show next, the structure of the extremal OTOC is rather rigid and only a very specific late-time correction to the density function  can change the OTOC (\ref{otoc:reg}) significantly for $t-t_{\rm eff}\gg \beta$.

Let us introduce a new time scale $t_\Lambda>t_*$ above which late-time corrections to the density function (\ref{densityfunction_late}) is significantly large.\footnote{Since the density function (\ref{rho:extremal}) is also exponentially suppressed for $t\ll t_{\rm eff}$, one might consider higher order $1/\N$ corrections to (\ref{rho:extremal}) in this regime. However, contributions of such early-time corrections to the OTOC are always small and hence they can never affect the general structure of the extremally chaotic OTOC. Whereas, late-time corrections are interesting, as they can change the asymptotic form of the OTOC in the limit $t\rightarrow \infty$.} We can make general comments about the effects of such late-time ($t>t_\Lambda$) corrections on  the asymptotic form of the OTOC. For simplicity, we approximate late-time corrections to the density function (\ref{rho:extremal}) that start becoming  important for $t>t_\Lambda$ as follows
\be\label{eq:tk}
\delta \rho(t) = \rho_\infty e^{-\frac{2 \pi a}{\beta}t}\Theta\(t-t_\Lambda\)\  .
\ee
The exponent  $a$ is constant but arbitrary, however it has a lower bound $a\ge 1$. This lower bound follows from the boundedness condition  of the density function (\ref{rho:bound}). Moreover, we have assumed that $t_\Lambda>t_*$, especially for $a=1$, such that parametrically $\delta \rho(t)<1/\N$. Our goal is to show that only a very specific $\delta \rho(t)$ can significantly change the late-time behavior of the OTOC. The above approximation (\ref{eq:tk}) for the late-time correction is sufficient to establish that point. We now argue that the density function can alter the asymptotic value of the OTOC only when it has a {\it long tail} that saturates the bound $a\ge1$ in (\ref{eq:tk}).  
 
 The contribution of a correction term (\ref{eq:tk}) to the OTOC can be computed analytically by using the representation (\ref{eq:KL}). At early times $t_\Lambda-t\gg \frac{\beta}{2\pi}$, we find that such contributions are exponentially suppressed 
 \be
 \delta F(t)=\frac{\rho_\infty \beta}{2\pi a} e^{\frac{2 \pi }{\beta}(t-t_*)}e^{\frac{2 \pi }{\beta}(t_*-a t_\Lambda)}+\cdots 
 \ee
for any $a\ge 1$. Likewise, the contribution of a correction term (\ref{eq:tk}) to the OTOC, for $a>1$,  decays exponentially fast even at late times $t-t_\Lambda\gg \frac{\beta}{2\pi}$
\be
 \delta F(t) \propto \beta\rho_\infty  e^{-\frac{2 \pi }{\beta}(t-t_\Lambda)}e^{-\frac{2 \pi }{\beta}(a-1) t_\Lambda}\ ,
\ee
implying the OTOC still asymptotes to $F=F_d$ in the limit $t\rightarrow \infty$. So, a correction term (\ref{eq:tk}) with $a>1$ can only control how the OTOC approaches its asymptotic value $F=F_d$, however, it cannot change the general structure of the extremal OTOC (\ref{eq:intro}).

 On the other hand, the same is not true when the density function has a long tail: $a=1$.\footnote{Note that now there is also a bound on $\rho_\infty$ from (\ref{rho:bound}): $\frac{8F_d}{\beta}\ge \rho_\infty\ge 0$.} In this case, we obtain 
 \be\label{eq:heavy}
 \delta F(t)=-\frac{\rho_\infty \beta}{4\pi }\(\pi -2 \tan ^{-1}\left(e^{\frac{2 \pi  (t_\Lambda-t)}{\beta }}\right)\)
 \ee
 implying that the full OTOC now asymptotes to
 \be
 F(t\gg t_\Lambda)=F_d-\frac{\rho_\infty \beta}{4 }\ ,
 \ee
 irrespective of the value of $t_\Lambda$. So, in this case the regularized OTOC (\ref{otoc:reg}) provides a good approximation only up to the cut-off scale $t_\Lambda$. 
 
 We wish to note that the connection between the asymptotic value of the OTOC and long tails of the density function is true in general. We will discuss this more in the next section.

 So, long-tailed corrections of (regularized or unregularized) extremally chaotic OTOCs are interesting since they can change the asymptotic structure of the extremal OTOC (\ref{eq:intro}).   Furthermore, we expect that such long-tailed corrections are generated in theories of quantum gravity from higher-order $1/\N$ effects. For example, the density functions do exhibit long tails in the Schwarzian theory (see appendix \ref{app:ST}) and also in 2d CFT with a large central charge.\footnote{This fact can be deduced in 2d CFT directly from \cite{Roberts:2014ifa}.} 
 
  \begin{figure}
\centering
\includegraphics[scale=0.5]{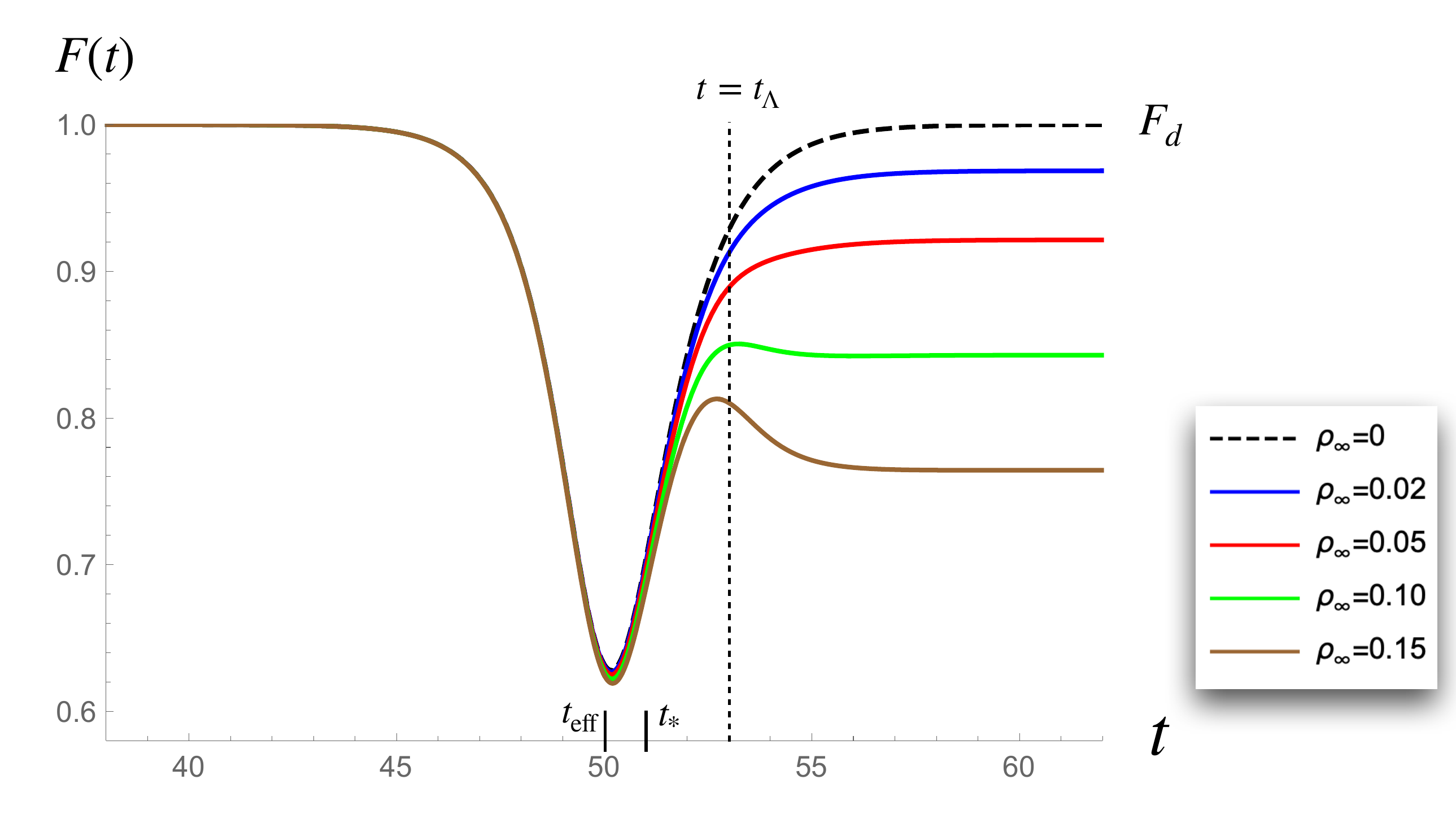}
\caption{ \label{figure:tail} \small  A long-tailed correction can change the asymptotic behavior of the extremally chaotic OTOC. This is a plot of the (regularized) extremally chaotic OTOC with long-tailed corrections $F_{\rm ext}^{\rm reg}(t)+ \delta F(t)$ for various values of $\rho_\infty$. For the plot, we have chosen $\beta=2\pi, F_d=1, c_1=2, t_{\rm eff}=50, \epsilon=0.5, t_*=51$ and $t_\Lambda=53$, where $t_\Lambda$ is the time scale at which long-tailed corrections start to become important. At early times $t_\Lambda-t\gg \frac{\beta}{2\pi}$, the OTOC remains unaffected. }
\end{figure}

It is, therefore, interesting that extremal OTOCs  with a long-tailed correction 
\be\label{QG}
F(t)=F_{\rm ext}^{\rm reg}(t)+ \delta F(t)\ ,
\ee
where $ \delta F(t)$ is given in (\ref{eq:heavy}),  have features that are qualitatively very similar (see figure \ref{figure:tail}) to the spectral form factor of the SYK model \cite{Cotler:2016fpe}.   This suggests that there is universality in the ramp-and-plateau behavior at very late times. Perhaps, a similar analysis can also be performed for the spectral form factor, though we will have to leave this question for the future.

 \section{Analytic Completions of Maximal Chaos}\label{sec:maximal}
 The $i\epsilon$-regularized OTOC (\ref{otoc:reg}), with or without a long-tailed correction, is an example of a small deformation of the extremally chaotic OTOC (\ref{eq:intro}). One advantage of the spectral representation (\ref{eq:KL}) is that it enables us to study general small deformation of the extremal OTOC (\ref{eq:intro}).

\subsection{Small Deformations of Extremal Chaos}
We begin by providing a precise definition of a general small deformation. We define small deformations as follows: the density function $\rho(t)$ is a narrow distribution (however can be a complicated function) which is small everywhere outside a window $t_{\rm eff}-\Delta t_{\rm eff} \le t \le t_{\rm eff} +\Delta t_{\rm eff}$, where $\Delta t_{\rm eff} \ll t_{\rm eff}$. In particular, it obeys 
\be\label{eq:narrow}
\int_{t_{\rm eff}-\Delta t_{\rm eff}}^{t_{\rm eff}+\Delta t_{\rm eff}} dt' \rho(t')\gg \int_{t_0}^{t_{\rm eff}-\Delta t_{\rm eff}} dt' \rho(t')+\int_{t_{\rm eff}+\Delta t_{\rm eff}}^\infty dt' \rho(t')\ .
\ee  
Moreover, the density function must also obey all the properties that follow from (\ref{eq:inversion}) as discussed before. The resulting OTOCs have the following properties. In the  regime $t_0\ll t\ll t_{\rm eff},t_*$, they saturate the MSS bound.  Furthermore, they are analytic functions obeying the Schwarz reflection (\ref{eq:SR}) and the boundedness (\ref{eq:positive}) conditions in the half-strip (\ref{halfstrip}) (including the boundary). However, these OTOCs, in general, saturate (exactly or approximately) only a subset of all the chaos bounds in any time duration. 

Any analytic completion of a long period of maximal chaos  (\ref{maximal}) must be a small deformation of extremal chaos. This becomes obvious once we rewrite  (\ref{eq:KL}) as
\be\label{eq:FD}
 F_d-F(t)=\frac{c_1}{\N} e^{\frac{2\pi}{\beta} t}-  e^{\frac{2\pi}{\beta} t}\int_{t_0}^\infty dt' f_{FD}(t;t')\rho(t')+\O\(e^{-\frac{2\pi (t-t_0)}{\beta}}\)\ ,
\ee
where $f_{FD}(t;t')\equiv 1- e^{-\frac{2\pi}{\beta} t} \mathcal{F}_{\rm ext}(t;t')$ and $\int_{t_0}^\infty dt' \rho(t')=\frac{c_1}{\N}$. The function $f_{FD}(t;t')$ is exactly the Fermi-Dirac distribution that asymptotes between $1$ for $t'\ll t$ and $0$ for $t'\gg t$. This provides some insight into the spectral representation of the OTOC. Let us now consider an OTOC which is maximally chaotic (\ref{maximal}) for  $t_0\ll  t\le t_i$. This necessarily requires that the second term in (\ref{eq:FD}) is small compared to the first term for $t\le t_i$, implying
\begin{align}
\int_{t_0}^\infty dt' \rho(t')\gg \int_{t_0}^\infty dt' f_{FD}(t_i;t')\rho(t')> \int_{t_0}^{t_i} dt' f_{FD}(t_i;t')\rho(t')\ .
\end{align}
The last integral has a two-sided bound
\be
\frac{1}{2}\int_{t_0}^{t_i} dt' \rho(t')\le \int_{t_0}^{t_i} dt' f_{FD}(t_i;t')\rho(t')\le \int_{t_0}^{t_i} dt' \rho(t')
\ee
that leads to 
\be\label{max_con1}
\int_{t_0}^\infty dt' \rho(t')\gg \int_{t_0}^{t_i} dt' \rho(t')\ .
\ee
Moreover, we assume that the OTOC starts to deviate from the maximally chaotic OTOC (\ref{maximal})  only near the scrambling time scale. In other words,   the difference between two time scales $t_i$ and $t_*$ is not too large compared to  $\frac{\beta}{2\pi}$.

 On the other hand, the bound (\ref{rho:bound}) dictates that the density function is also small $\rho(t)\ll 1/\N$ for $t\ge t_f$, where $t_f=t_*+\O(1) \frac{\beta}{2\pi}$. In particular, for any such $t_f$
 \be
 \int^{\infty}_{t_f} dt' \rho(t')<\frac{4 F_d}{\pi}e^{-\frac{2\pi}{\beta} t_f}\ll \frac{1}{\N}\ .
 \ee
Hence, the above integral is exponentially suppressed compared to $\int_{t_0}^\infty dt \rho(t)=\frac{c_1}{\N}$, which is the coefficient of the term $e^{\frac{2\pi t}{\beta}}$ in (\ref{maximal}).\footnote{Throughout the paper, we are assuming that $F_d$ is order 1. This can be ensured by normalizing operators $V$ and $W$ appropriately. } This immediately implies (\ref{eq:narrow}) with $t_i\le t_{\rm eff} \le t_f$ and $\Delta t_{\rm eff}\le \frac{1}{2}(t_f-t_i)\sim \O(1) \frac{\beta}{2\pi}$. Note that $t_{\rm eff}$ and $t_*$ are not parametrically separated when we have a long period of maximal chaos. So, $t_{\rm eff}$ can simply be $t_*$ or it can also be an independent time scale. In any case, for strongly chaotic systems $F_d-F(t\sim t_*)\sim \O(1) F_d$, indicating that the difference between these two time scales is not very large compared to  $\frac{\beta}{2\pi}$.

 So, we conclude that all analytic completions of a long period of maximal chaos  are small deformations of extremal chaos.\footnote{The discussion of this section is valid even when the period of maximal chaos is short, \ie $t_i\ll t_*$. For such systems, the only difference is that $\Delta t_{\rm eff}$ can be large $\Delta t_{\rm eff}\gg \frac{\beta}{2\pi}$.} It should be noted that the resulting OTOCs can defer significantly from the extremal OTOC (\ref{eq:intro}), however, from the perspective of the density function they are always small deformations. As a consequence, these OTOCs have universal qualitative features far away from $t=t_{\rm eff}$, as we discuss next.

\subsection{Early-Time Behavior}
We first consider the regime $t\gg t_0$ and $t_{\rm eff}-t\gg \frac{\beta}{2\pi}$. The integral (\ref{eq:FD}) in this limit can be simplified, obtaining
\be\label{OTOC:early}
F(t)=F_d-e^{\frac{2\pi t}{\beta}}\(\frac{c_1}{\N}-\int_{t_0}^{t_{\rm eff}-\Delta t_{\rm eff}} f_{FD}(t;t')\rho(t')dt' -c_2 e^{\frac{4\pi}{\beta} (t-t_{\rm eff})} \)+\cdots\ ,
\ee
where dots represent terms that are further exponentially suppressed $\O(e^{\frac{8\pi}{\beta} (t-t_{\rm eff})}, e^{\frac{2\pi}{\beta} (t_0-t)})$. Note that $\rho(t)$ is vanishingly small near $t=t_0$ and hence the above expression is independent of $t_0$. Moreover,  $c_1$ and $c_2$ coefficients are given by 
\be
\frac{c_1}{\N}=\int_{t_0}^\infty \rho(t')dt'\ , \qquad c_2\approx \int_{t_{\rm eff}-\Delta t_{\rm eff}}^{t_{\rm eff}+\Delta t_{\rm eff}} \rho(t')e^{\frac{4\pi}{\beta} (t_{\rm eff}-t')}dt' \ .
\ee
Both corrections to the maximally chaotic term in (\ref{OTOC:early}) are such that they slow down the initial $e^{\frac{2\pi t}{\beta}}$ growth of the OTOC.\footnote{Analyticity of $F(t)$ dictates that $\rho(t)$ cannot be exactly zero outside $t_{\rm eff}-\Delta t_{\rm eff} \le t \le t_{\rm eff} +\Delta t_{\rm eff}$. So, the integral in (\ref{OTOC:early}) is small but non-zero. Of course, this contribution still has a fixed sign since the density function is always non-negative.} This is true irrespective of the details of the density function $\rho(t)$, provided $t_*,t_{\rm eff}\gg t_0$.

The expression (\ref{OTOC:early}) provides a physical interpretation of $\rho(t)$ in the early-time regime: $t\gg t_0$ and $t_{\rm eff}-t\gg \beta/2\pi$. If $\rho(t)$, in this regime, is a slowly varying function of time, one can use  the Sommerfeld approximation trick to simplify
\be
\int_{t_0}^{t_{\rm eff}-\Delta t_{\rm eff}} f_{FD}(t;t')\rho(t')dt' \approx \int_{t_0}^t dt' \rho (t')\ .
\ee
 One can interpret this subleading contribution in (\ref{OTOC:early}) as coming from a small correction to the Lyapunov exponent: $F_d-F(t) \sim \exp(\frac{2\pi}{\beta}t+\int dt \delta\lambda_L(t))$. Hence,  at the leading order we obtain 
\be\label{eq:epsilon}
\delta\lambda_L(t)=-\frac{\rho(t)}{\int_{t_0}^\infty \rho(t')dt'}\ ,
\ee
implying that the Lyapunov exponent decreases monotonically from the MSS saturation value as one approaches the time scale $t_{\rm eff}$.

It is well-known that holographic theories dual to Einstein gravity saturate the MSS bound, where $\frac{1}{\N}=\exp(\frac{2\pi}{\beta}t_*)$ is determined by the Newton constant $G_N$. However, if we include stringy correction to the Einstein gravity result of  $\lambda_L=\frac{2\pi}{\beta}$, the OTOC no longer saturates the MSS bound \cite{Shenker:2014cwa}. A comparison with the above result (\ref{eq:epsilon}) suggests that at early times $\rho(t)=\rho_0\approx$ constant, which is determined by the string scale. In particular, from \cite{Shenker:2014cwa} we find that for theories dual to planar AdS$_{d+1}$ black holes   
\be
\rho_0= \frac{\pi d(d-1)}{2\beta}\(\frac{l_s}{l_{\rm AdS}}\)^2\frac{c_1}{\N}\ ,
\ee
where $l_s$ is the string length and $l_{\rm AdS}$ is the AdS radius.

 \subsection{Late-Time Behavior}
We can perform a similar analysis at late times $t\gg t_{\rm eff}$, obtaining
\begin{align}\label{OTOC:late}
F(t)=F_d&-\frac{\beta}{4}\rho_\infty \nonumber\\
&-\int_{t_{\rm eff}+\Delta t_{\rm eff}}^\infty \frac{dt' }{1+e^{\frac{4\pi }{\beta}(t-t')}}\(\rho(t')e^{\frac{2\pi t}{\beta}}-\rho_\infty\)-\tilde{c}_2 e^{\frac{2\pi }{\beta}(t_{\rm eff}-t)}+\cdots\ ,
\end{align}
 where $\rho_\infty=\lim_{t\rightarrow \infty}e^{\frac{2\pi t}{\beta}}\rho(t)\ge 0$ and dots represent subleading terms. The coefficient $\tilde{c}_2$ is given by
 \be
 \tilde{c}_2=\int_{t_0}^{t_{\rm eff}+\Delta t_{\rm eff}} dt'  e^{\frac{4\pi }{\beta}(t'-t_{\rm eff})}\rho(t')\approx \int_{t_{\rm eff}-\Delta t_{\rm eff}}^{t_{\rm eff}+\Delta t_{\rm eff}} dt'  e^{\frac{4\pi }{\beta}(t'-t_{\rm eff})}\rho(t')
 \ee
which is strictly positive. Clearly, the first line of (\ref{OTOC:late}) is the asymptotic value of the OTOC for large $t$. This asymptotic value depends entirely on whether the density function has a long tail. The second line of (\ref{OTOC:late}) contains the leading terms that control how the OTOC approaches its asymptotic value. Note that the second line of (\ref{OTOC:late}) does not have a fixed sign in general since  the integral in (\ref{OTOC:late}) can have either sign when $\rho_\infty>0$.

~\\

Finally, we make some general comments. As stated before, the regularized extremally chaotic OTOC (\ref{otoc:reg}) can be regarded as a ``tree-level" analytic completion of the leading Lyapunov behavior (\ref{maximal}), where $e^{\frac{2\pi t_{\rm eff}}{\beta}}$ plays the role of an effective string scale. However, it should be noted that small deformations of extremal chaos, as described above, are more general. For example, they also capture analytic completions of the leading Lyapunov behavior (\ref{maximal}) by summing over higher order $1/\N$ contributions. In such a case, scales $t_*$ and $t_{\rm eff}$ are not independent since they are both determined by $\N$. This is exactly what happens in the Schwarzian theory, which describes 2D Jackiw-Teitelboim (JT) gravity \cite{Jackiw:1984je,Teitelboim:1983ux}, as we discuss in appendix \ref{app:ST}. In the Schwarzian theory, it is possible to compute the OTOC (\ref{eq:otoc}) as an expansion in $\beta/C = 1/\N$ (see \cite{Maldacena:2016upp,Lam:2018pvp}) that is analytic even in the regime $t>t_*$ where it asymptotes to zero. As a check, one can compute the associated density function which is indeed a  narrow distribution (a single peak with width $\sim \beta$) around $t\sim t_*$. Likewise, we observe from \cite{Roberts:2014ifa} that 2d CFTs with a large central charge $c$ have the same qualitative features when we include $1/c$ corrections. This is perfectly consistent with our general discussion that any analytic completion of maximal chaos must be a small deformation of extremal chaos. Moreover, the density functions for $t> t_*$, in both of these cases, have long tails with $\rho_\infty=\frac{4F_d}{\beta}$.

 \section{Conclusions and Outlook}\label{sec:conclusions} 
It is always important to ask what general lessons we can learn from various models of quantum gravity. Quantum chaos provides a profound answer to this question. There are compelling reasons to believe that all theories of quantum gravity and their holographic duals are maximally chaotic. More precisely, in these theories, the OTOC (\ref{eq:otoc}) saturates the MSS bound on chaos (\ref{MSS:Lyapunov}) for $t_0\ll t\ll t_*$. In this paper, we showed that  the extremally chaotic OTOC (\ref{eq:intro}) analytically completes the maximally chaotic OTOC inside the half-strip $\{t\in \mathbb{C}|\ \mbox{Re}\ t\gg t_0\ \text{and}\ |\mbox{Im}\ t|< \frac{\beta}{4} \}$, saturating even the subleading chaos bounds of \cite{Kundu:2021qcx}. Furthermore, we argued for its uniqueness. Interestingly, the extremal OTOC provides a spectral representation (\ref{intro:KL}) of all OTOCs that are analytic functions obeying the boundedness and the Schwarz reflection conditions in the half-strip (\ref{halfstrip}). A non-trivial implication of this representation is that all analytic completions of a long period of maximal chaos must be small deformations of extremal chaos from the perspective of the density function. 
 
Any physical system cannot be exactly extremally chaotic since the extremal OTOC (\ref{eq:intro}) is non-analytic on the boundary of the half-strip (\ref{halfstrip}). This problem can be resolved naturally by a standard $i\epsilon$-regularization, making the OTOC analytic everywhere in the half-strip (\ref{halfstrip}) obeying the Schwarz reflection (\ref{eq:SR}) and the boundedness (\ref{eq:positive}) conditions. The regularized OTOC, for real $t$, has the same qualitative features as the extremal OTOC, differing only slightly for $t>t_{\rm eff}$ even when $\epsilon$ is finite. Thus, a physical system, in principle, can be approximately extremally chaotic up to some time scale $t_\Lambda>t_*$.\footnote{Of course, it is completely self consistent to have $t_\Lambda=\infty$. } Unfortunately, we are not aware of any chaotic system which exhibits this behavior even qualitatively. It would be interesting to construct such systems.
  
At this point, it is reasonable to make the conjecture that all theories of quantum gravity (and their holographic duals) are strongly chaotic systems that are small deformations of extremally chaotic systems. In particular, the associated density function $\rho(t)$ is a narrow distribution  that is small everywhere outside a window around $t_{\rm eff}$ obeying (\ref{eq:narrow}). Of course, $\rho(t)$ can be a complicated function inside the narrow window. Moreover,  the integral $\int_{t_0}^\infty dt \rho(t)=\frac{c_1}{\N}$ is exactly the coefficient of the term $e^{\frac{2\pi t}{\beta}}$ in (\ref{maximal}), determining the scrambling time scale $t_*$. This conjecture follows directly from the fast scrambling conjecture of \cite{Sekino:2008he} (or equivalently the more precise version in terms of the MSS bound \cite{Maldacena:2015waa}).  

A more tempting but speculative conjecture would be that OTOCs associated with quantum gravity have qualitative features similar to  the extremally chaotic OTOC with a long-tailed correction (\ref{QG}). This conjecture perhaps can be checked in the SYK model by utilizing numerical tools developed in \cite{Kobrin:2020xms}.

\section*{Acknowledgments}

It is my pleasure to thank Diptarka Das, Thomas Hartman, Jared Kaplan,  Arnab Kundu, Eric Perlmutter, and Douglas Stanford  for helpful discussions and commenting on a draft. I was supported in part by the Simons Collaboration Grant on the Non-Perturbative Bootstrap.

\begin{appendix}
\section{Schwarzian Theory and Late-Time Long Tail}\label{app:ST}
The Sachdev-Ye-Kitaev (SYK) model is maximally chaotic \cite{kitaev2014hidden,Sachdev:1992fk,Polchinski:2016xgd,Maldacena:2016hyu,Jevicki:2016bwu,Jevicki:2016ito,Cotler:2016fpe}. The SYK model also exhibits other features suggestive of the fact that it is the holographic dual of a 2D dilaton gravity theory on AdS.  It was pointed out by Kitaev \cite{kitaev2014hidden,Kitaev:2017awl} that the IR dynamics of the SYK model is described by the Schwarzian theory of a single effective degree of freedom $t(u)$
 \be
 S=-C\int du \{t(u),u\}\ ,
 \ee
where the Schwarzian derivative $\{t(u),u\}=\frac{t'''}{t'}-\frac{3}{2}\frac{t''^2}{t'^2}$. The Schwarzian theory also describes JT gravity \cite{Jackiw:1984je,Teitelboim:1983ux}, providing a precise connection between the SYK model and 2D dilaton gravity theory \cite{Almheiri:2014cka,Jensen:2016pah,Maldacena:2016upp,Engelsoy:2016xyb,Cvetic:2016eiv,Nayak:2018qej}. 

In the Schwarzian theory, it is possible to compute the OTOC (\ref{eq:otoc}) as an expansion in $\beta/C\equiv 1/\N$. In particular, the OTOC is given by  the confluent hypergeometric $U$-function \cite{Maldacena:2016upp,Lam:2018pvp}
\be\label{otoc:sch}
\frac{F(t)}{F_d}=\frac{1}{z^{2\Delta}}U\(2\Delta, 1, \frac{1}{z}\)\ , \qquad z=\frac{\beta}{16\pi C}e^{\frac{2\pi t}{\beta}}\equiv \frac{1}{16\pi }e^{\frac{2\pi }{\beta}(t-t_*)}
\ee
in the limit $C/\beta\gg 1$, $e^{\frac{2\pi t}{\beta}}\gg 1$ with $z$ fixed, where operators $V$ and $W$ have the same scaling dimension $\Delta$. This OTOC is maximally chaotic for $t_*-t\gg \frac{\beta}{2\pi}$. Moreover, it is also well-behaved for $t>t_*$, where it asymptotes to zero. This is in sharp contrast with the extremally chaotic OTOC (\ref{eq:intro}), which asymptotes to $F_d$ for $t>t_*$.

In the language of extremal chaos,  the OTOC (\ref{otoc:sch}) is a special case in which two time scales: $t_*, t_{\rm eff}$ are not independent. Nevertheless, the discussion of this paper still applies implying that the OTOC (\ref{otoc:sch}) is a small deformation of the extremal OTOC. Besides, we will show that the OTOC (\ref{otoc:sch}) has a density function with a long tail. 

\begin{figure}
\centering
\includegraphics[scale=0.5]{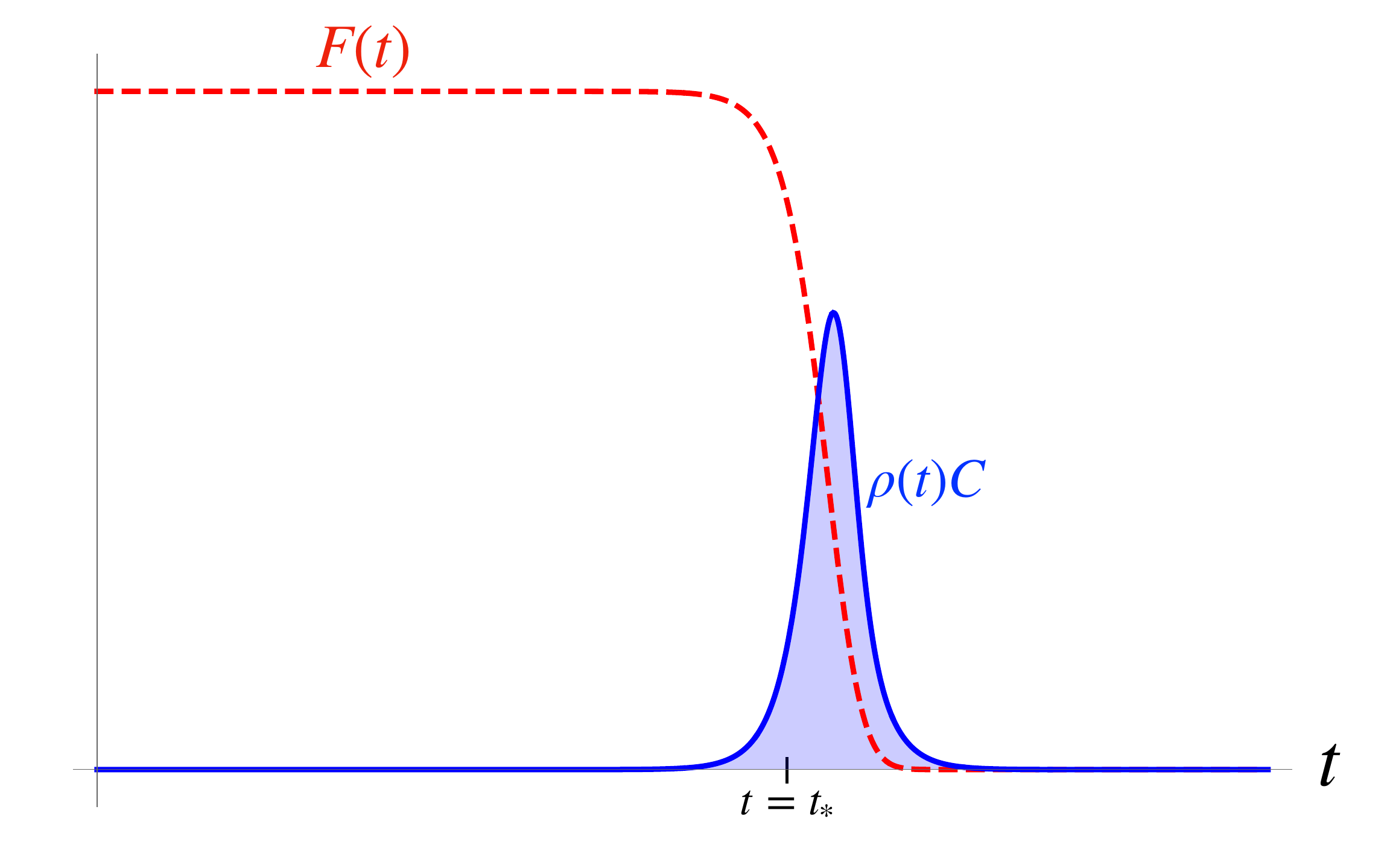}
\caption{ \label{contour} \small  A generic OTOC and the associated density function (normalized by $C$) are shown for the Schwarzian theory. The density function $\rho(t)$ has a narrow peak of width $\sim \beta$ near the scrambling time $t_*=\frac{\beta}{2\pi}\ln (C/\beta)$. So from the perspective of the density function, this is exactly a small deformation of extremal chaos where the time scale $t_{\rm eff}$ is also set by the scrambling time. Note that OTOCs of 2d CFTs with a large central charge also have the same qualitative features. }\label{fig:sch}
\end{figure}

We begin by noting the asymptotic behaviors of the OTOC (\ref{otoc:sch}). Initially, the leading growing term saturates the MSS bound
\be
F(t)=F_d \(1-\frac{  \Delta ^2 }{4 \pi }e^{\frac{2 \pi  (t-t_*)}{\beta }}+\frac{ (2 \Delta +1)^2 \Delta ^2 }{128 \pi ^2}e^{\frac{4 \pi  (t-t_*)}{\beta }}+\cdots\)\qquad  \frac{\beta}{2\pi}\ll t, t_*-t
\ee
however, correction terms start to become important near the scrambling time. On the other hand, $F(t)$ decays exponentially at late times
\be
F(t)=\frac{2\pi F_d}{\beta} \frac{(16\pi)^{2\Delta}}{\Gamma(2\Delta)}(t-t_*)e^{-\frac{4 \pi \Delta  (t-t_*)}{\beta }}+\cdots \qquad  t- t_*\gg \frac{\beta}{2\pi}\ .
\ee
We can now obtain the density function $\rho(t)$ by using the inversion formula (\ref{eq:inversion}):
 \be
 \rho(t)=\frac{4F_d}{\beta}e^{-\frac{2\pi t}{\beta}}\(1-\mbox{Re}\ \frac{e^{-i \pi \Delta}U\(2\Delta, 1,- \frac{i}{z}\)}{z^{2\Delta}}\) \qquad z= \frac{1}{16\pi }e^{\frac{2\pi }{\beta}(t-t_*)}
 \ee
 which is a narrow distribution around $t\sim t_*$, as shown in figure \ref{fig:sch}. In particular, the density function is small away from the peak
 \begin{align}
\rho(t)=F_d\frac{\Delta ^2 (2 \Delta +1)^2 }{32 \pi ^2 C}e^{\frac{2 \pi  (t-t_*)}{\beta }}\ , \qquad \qquad  \frac{\beta}{2\pi}\ll t, t_*-t\ .
 \end{align}
Similarly at late times $t- t_*\gg \frac{\beta}{2\pi}$, the density function is given by 
 \be
 \rho(t)=\frac{4 F_d}{\beta}e^{-\frac{2\pi t}{\beta}}-\frac{8\pi F_d}{\beta^2} \frac{(16\pi)^{2\Delta}}{\Gamma(2\Delta)}e^{-\frac{2\pi t}{\beta}-\frac{4 \pi \Delta  (t-t_*)}{\beta }}\((t-t_*)\cos(\pi \Delta)+\frac{\beta}{4}\sin(\pi \Delta)\)
 \ee
implying that the density function has a long tail. Note that $C \rho(t)/\beta$, for $t\gg t_*$, is exponentially suppressed. So, the long tail of $\rho(t)$, strictly speaking, is not a perturbative  $\frac{1}{C}$ effect.

 All these results are perfectly consistent with our general discussion that any analytic completion of maximal chaos must be a small deformation of extremal chaos. For example, one can check that the above asymptotic expressions satisfy (\ref{OTOC:early}) and (\ref{OTOC:late}). However, the Schwarzian theory represents a very special case in which $F(t)$ goes to zero for $t\gg t_*$ indicating that the information about the initial perturbation is completely lost.  Of course, the OTOC (\ref{otoc:sch}) is not valid in the limit $t\rightarrow \infty$. It is definitely possible that non-perturbative contributions to $F(t)$ are such that the exact OTOC asymptotes to some non-zero value in the limit $t\rightarrow \infty$. 
 
A similar analysis can also be performed for  2d CFTs with a large central charge $c$, where we also know $F(t)$ beyond the leading $\frac{1}{c}$ term \cite{Roberts:2014ifa}. The corresponding density function exhibits very similar features with exactly the same long tail. 

\end{appendix}


\end{spacing}

\bibliographystyle{utphys} 
\bibliography{chaos}

\providecommand{\href}[2]{#2}\begingroup\raggedright\begin{thebibliography}{10}

\bibitem{Maldacena:2015waa}
J.~Maldacena, S.~H. Shenker, and D.~Stanford, ``{A bound on chaos},''
\href{http://arxiv.org/abs/1503.01409}{{\ttfamily arXiv:1503.01409 [hep-th]}}.

\bibitem{Sekino:2008he}
Y.~Sekino and L.~Susskind, ``{Fast Scramblers},''
  \href{http://dx.doi.org/10.1088/1126-6708/2008/10/065}{{\em JHEP} {\bfseries
  10} (2008) 065}, \href{http://arxiv.org/abs/0808.2096}{{\ttfamily
  arXiv:0808.2096 [hep-th]}}.

\bibitem{Kundu:2021qcx}
S.~Kundu, ``{Subleading Bounds on Chaos},''
  \href{http://arxiv.org/abs/2109.03826}{{\ttfamily arXiv:2109.03826
  [hep-th]}}.

\bibitem{Poojary:2018esz}
R.~R. Poojary, ``{BTZ dynamics and chaos},''
  \href{http://dx.doi.org/10.1007/JHEP03(2020)048}{{\em JHEP} {\bfseries 03}
  (2020) 048}, \href{http://arxiv.org/abs/1812.10073}{{\ttfamily
  arXiv:1812.10073 [hep-th]}}.

\bibitem{Banerjee:2019vff}
A.~Banerjee, A.~Kundu, and R.~R. Poojary, ``{Rotating black holes in AdS
  spacetime, extremality, and chaos},''
  \href{http://dx.doi.org/10.1103/PhysRevD.102.106013}{{\em Phys. Rev. D}
  {\bfseries 102} no.~10, (2020) 106013},
  \href{http://arxiv.org/abs/1912.12996}{{\ttfamily arXiv:1912.12996
  [hep-th]}}.

\bibitem{Craps:2020ahu}
B.~Craps, M.~De~Clerck, P.~Hacker, K.~Nguyen, and C.~Rabideau, ``{Slow
  scrambling in extremal BTZ and microstate geometries},''
  \href{http://dx.doi.org/10.1007/JHEP03(2021)020}{{\em JHEP} {\bfseries 03}
  (2021) 020}, \href{http://arxiv.org/abs/2009.08518}{{\ttfamily
  arXiv:2009.08518 [hep-th]}}.

\bibitem{Craps:2021bmz}
B.~Craps, S.~Khetrapal, and C.~Rabideau, ``{Chaos in CFT dual to rotating
  BTZ},'' \href{http://arxiv.org/abs/2107.13874}{{\ttfamily arXiv:2107.13874
  [hep-th]}}.

\bibitem{Roberts:2014ifa}
D.~A. Roberts and D.~Stanford, ``{Two-dimensional conformal field theory and
  the butterfly effect},''
\href{http://arxiv.org/abs/1412.5123}{{\ttfamily arXiv:1412.5123 [hep-th]}}.

\bibitem{Shenker:2014cwa}
S.~H. Shenker and D.~Stanford, ``{Stringy effects in scrambling},''
  \href{http://dx.doi.org/10.1007/JHEP05(2015)132}{{\em JHEP} {\bfseries 05}
  (2015) 132}, \href{http://arxiv.org/abs/1412.6087}{{\ttfamily arXiv:1412.6087
  [hep-th]}}.

\bibitem{Vladimirov}
V.~Vladimirov, {\em {Methods of the theory of functions of many complex
  variables}}.
\newblock MIT Press, Cambridge, U.S.A.

\bibitem{Kravchuk:2020scc}
P.~Kravchuk, J.~Qiao, and S.~Rychkov, ``{Distributions in CFT. Part I.
  Cross-ratio space},'' \href{http://dx.doi.org/10.1007/JHEP05(2020)137}{{\em
  JHEP} {\bfseries 05} (2020) 137},
  \href{http://arxiv.org/abs/2001.08778}{{\ttfamily arXiv:2001.08778
  [hep-th]}}.

\bibitem{Blake:2017ris}
M.~Blake, H.~Lee, and H.~Liu, ``{A quantum hydrodynamical description for
  scrambling and many-body chaos},''
  \href{http://dx.doi.org/10.1007/JHEP10(2018)127}{{\em JHEP} {\bfseries 10}
  (2018) 127}, \href{http://arxiv.org/abs/1801.00010}{{\ttfamily
  arXiv:1801.00010 [hep-th]}}.

\bibitem{Blake:2021wqj}
M.~Blake and H.~Liu, ``{On systems of maximal quantum chaos},''
  \href{http://dx.doi.org/10.1007/JHEP05(2021)229}{{\em JHEP} {\bfseries 05}
  (2021) 229}, \href{http://arxiv.org/abs/2102.11294}{{\ttfamily
  arXiv:2102.11294 [hep-th]}}.

\bibitem{Grozdanov:2017ajz}
S.~Grozdanov, K.~Schalm, and V.~Scopelliti, ``{Black hole scrambling from
  hydrodynamics},''
  \href{http://dx.doi.org/10.1103/PhysRevLett.120.231601}{{\em Phys. Rev.
  Lett.} {\bfseries 120} no.~23, (2018) 231601},
  \href{http://arxiv.org/abs/1710.00921}{{\ttfamily arXiv:1710.00921
  [hep-th]}}.

\bibitem{Haehl:2018izb}
F.~M. Haehl and M.~Rozali, ``{Effective Field Theory for Chaotic CFTs},''
  \href{http://dx.doi.org/10.1007/JHEP10(2018)118}{{\em JHEP} {\bfseries 10}
  (2018) 118}, \href{http://arxiv.org/abs/1808.02898}{{\ttfamily
  arXiv:1808.02898 [hep-th]}}.

\bibitem{Blake:2018leo}
M.~Blake, R.~A. Davison, S.~Grozdanov, and H.~Liu, ``{Many-body chaos and
  energy dynamics in holography},''
  \href{http://dx.doi.org/10.1007/JHEP10(2018)035}{{\em JHEP} {\bfseries 10}
  (2018) 035}, \href{http://arxiv.org/abs/1809.01169}{{\ttfamily
  arXiv:1809.01169 [hep-th]}}.

\bibitem{Grozdanov:2018kkt}
S.~Grozdanov, ``{On the connection between hydrodynamics and quantum chaos in
  holographic theories with stringy corrections},''
  \href{http://dx.doi.org/10.1007/JHEP01(2019)048}{{\em JHEP} {\bfseries 01}
  (2019) 048}, \href{http://arxiv.org/abs/1811.09641}{{\ttfamily
  arXiv:1811.09641 [hep-th]}}.

\bibitem{Haehl:2019eae}
F.~M. Haehl, W.~Reeves, and M.~Rozali, ``{Reparametrization modes, shadow
  operators, and quantum chaos in higher-dimensional CFTs},''
  \href{http://dx.doi.org/10.1007/JHEP11(2019)102}{{\em JHEP} {\bfseries 11}
  (2019) 102}, \href{http://arxiv.org/abs/1909.05847}{{\ttfamily
  arXiv:1909.05847 [hep-th]}}.

\bibitem{Ahn:2019rnq}
Y.~Ahn, V.~Jahnke, H.-S. Jeong, and K.-Y. Kim, ``{Scrambling in Hyperbolic
  Black Holes: shock waves and pole-skipping},''
  \href{http://dx.doi.org/10.1007/JHEP10(2019)257}{{\em JHEP} {\bfseries 10}
  (2019) 257}, \href{http://arxiv.org/abs/1907.08030}{{\ttfamily
  arXiv:1907.08030 [hep-th]}}.

\bibitem{Ahn:2020bks}
Y.~Ahn, V.~Jahnke, H.-S. Jeong, K.-Y. Kim, K.-S. Lee, and M.~Nishida,
  ``{Pole-skipping of scalar and vector fields in hyperbolic space: conformal
  blocks and holography},''
  \href{http://dx.doi.org/10.1007/JHEP09(2020)111}{{\em JHEP} {\bfseries 09}
  (2020) 111}, \href{http://arxiv.org/abs/2006.00974}{{\ttfamily
  arXiv:2006.00974 [hep-th]}}.

\bibitem{Ramirez:2020qer}
D.~M. Ramirez, ``{Chaos and pole skipping in CFT$_2$},''
  \href{http://arxiv.org/abs/2009.00500}{{\ttfamily arXiv:2009.00500
  [hep-th]}}.

\bibitem{Choi:2020tdj}
C.~Choi, M.~Mezei, and G.~S\'arosi, ``{Pole skipping away from maximal
  chaos},'' \href{http://arxiv.org/abs/2010.08558}{{\ttfamily arXiv:2010.08558
  [hep-th]}}.

\bibitem{Roberts:2014isa}
D.~A. Roberts, D.~Stanford, and L.~Susskind, ``{Localized shocks},''
  \href{http://dx.doi.org/10.1007/JHEP03(2015)051}{{\em JHEP} {\bfseries 03}
  (2015) 051}, \href{http://arxiv.org/abs/1409.8180}{{\ttfamily arXiv:1409.8180
  [hep-th]}}.

\bibitem{Shenker:2013pqa}
S.~H. Shenker and D.~Stanford, ``{Black holes and the butterfly effect},''
  \href{http://dx.doi.org/10.1007/JHEP03(2014)067}{{\em JHEP} {\bfseries 03}
  (2014) 067},
\href{http://arxiv.org/abs/1306.0622}{{\ttfamily arXiv:1306.0622 [hep-th]}}.

\bibitem{Shenker:2013yza}
S.~H. Shenker and D.~Stanford, ``{Multiple Shocks},''
  \href{http://dx.doi.org/10.1007/JHEP12(2014)046}{{\em JHEP} {\bfseries 12}
  (2014) 046}, \href{http://arxiv.org/abs/1312.3296}{{\ttfamily arXiv:1312.3296
  [hep-th]}}.

\bibitem{kitaev2014hidden}
A.~Kitaev, ``Hidden correlations in the hawking radiation and thermal noise,''
  in {\em https://youtu.be/OQ9qN8j7EZI}.

\bibitem{Kundu:2020gkz}
S.~Kundu, ``{A Generalized Nachtmann Theorem in CFT},''
  \href{http://dx.doi.org/10.1007/JHEP11(2020)138}{{\em JHEP} {\bfseries 11}
  (2020) 138}, \href{http://arxiv.org/abs/2002.12390}{{\ttfamily
  arXiv:2002.12390 [hep-th]}}.

\bibitem{Kundu:2021qpi}
S.~Kundu, ``{Swampland Conditions for Higher Derivative Couplings from CFT},''
  \href{http://arxiv.org/abs/2104.11238}{{\ttfamily arXiv:2104.11238
  [hep-th]}}.

\bibitem{Maldacena:2016upp}
J.~Maldacena, D.~Stanford, and Z.~Yang, ``{Conformal symmetry and its breaking
  in two dimensional Nearly Anti-de-Sitter space},''
  \href{http://dx.doi.org/10.1093/ptep/ptw124}{{\em PTEP} {\bfseries 2016}
  no.~12, (2016) 12C104}, \href{http://arxiv.org/abs/1606.01857}{{\ttfamily
  arXiv:1606.01857 [hep-th]}}.

\bibitem{Lam:2018pvp}
H.~T. Lam, T.~G. Mertens, G.~J. Turiaci, and H.~Verlinde, ``{Shockwave S-matrix
  from Schwarzian Quantum Mechanics},''
  \href{http://dx.doi.org/10.1007/JHEP11(2018)182}{{\em JHEP} {\bfseries 11}
  (2018) 182}, \href{http://arxiv.org/abs/1804.09834}{{\ttfamily
  arXiv:1804.09834 [hep-th]}}.

\bibitem{Cotler:2016fpe}
J.~S. Cotler, G.~Gur-Ari, M.~Hanada, J.~Polchinski, P.~Saad, S.~H. Shenker,
  D.~Stanford, A.~Streicher, and M.~Tezuka, ``{Black Holes and Random
  Matrices},''
\href{http://arxiv.org/abs/1611.04650}{{\ttfamily arXiv:1611.04650 [hep-th]}}.

\bibitem{Jackiw:1984je}
R.~Jackiw, ``{Lower Dimensional Gravity},''
  \href{http://dx.doi.org/10.1016/0550-3213(85)90448-1}{{\em Nucl. Phys. B}
  {\bfseries 252} (1985) 343--356}.

\bibitem{Teitelboim:1983ux}
C.~Teitelboim, ``{Gravitation and Hamiltonian Structure in Two Space-Time
  Dimensions},'' \href{http://dx.doi.org/10.1016/0370-2693(83)90012-6}{{\em
  Phys. Lett. B} {\bfseries 126} (1983) 41--45}.

\bibitem{Kobrin:2020xms}
B.~Kobrin, Z.~Yang, G.~D. Kahanamoku-Meyer, C.~T. Olund, J.~E. Moore,
  D.~Stanford, and N.~Y. Yao, ``{Many-Body Chaos in the Sachdev-Ye-Kitaev
  Model},'' \href{http://dx.doi.org/10.1103/PhysRevLett.126.030602}{{\em Phys.
  Rev. Lett.} {\bfseries 126} no.~3, (2021) 030602},
  \href{http://arxiv.org/abs/2002.05725}{{\ttfamily arXiv:2002.05725
  [hep-th]}}.

\bibitem{Sachdev:1992fk}
S.~Sachdev and J.~Ye, ``{Gapless spin fluid ground state in a random, quantum
  Heisenberg magnet},''
  \href{http://dx.doi.org/10.1103/PhysRevLett.70.3339}{{\em Phys. Rev. Lett.}
  {\bfseries 70} (1993) 3339},
  \href{http://arxiv.org/abs/cond-mat/9212030}{{\ttfamily
  arXiv:cond-mat/9212030}}.

\bibitem{Polchinski:2016xgd}
J.~Polchinski and V.~Rosenhaus, ``{The Spectrum in the Sachdev-Ye-Kitaev
  Model},'' \href{http://dx.doi.org/10.1007/JHEP04(2016)001}{{\em JHEP}
  {\bfseries 04} (2016) 001}, \href{http://arxiv.org/abs/1601.06768}{{\ttfamily
  arXiv:1601.06768 [hep-th]}}.

\bibitem{Maldacena:2016hyu}
J.~Maldacena and D.~Stanford, ``{Remarks on the Sachdev-Ye-Kitaev model},''
  \href{http://dx.doi.org/10.1103/PhysRevD.94.106002}{{\em Phys. Rev. D}
  {\bfseries 94} no.~10, (2016) 106002},
  \href{http://arxiv.org/abs/1604.07818}{{\ttfamily arXiv:1604.07818
  [hep-th]}}.

\bibitem{Jevicki:2016bwu}
A.~Jevicki, K.~Suzuki, and J.~Yoon, ``{Bi-Local Holography in the SYK Model},''
  \href{http://dx.doi.org/10.1007/JHEP07(2016)007}{{\em JHEP} {\bfseries 07}
  (2016) 007}, \href{http://arxiv.org/abs/1603.06246}{{\ttfamily
  arXiv:1603.06246 [hep-th]}}.

\bibitem{Jevicki:2016ito}
A.~Jevicki and K.~Suzuki, ``{Bi-Local Holography in the SYK Model:
  Perturbations},'' \href{http://dx.doi.org/10.1007/JHEP11(2016)046}{{\em JHEP}
  {\bfseries 11} (2016) 046}, \href{http://arxiv.org/abs/1608.07567}{{\ttfamily
  arXiv:1608.07567 [hep-th]}}.

\bibitem{Kitaev:2017awl}
A.~Kitaev and S.~J. Suh, ``{The soft mode in the Sachdev-Ye-Kitaev model and
  its gravity dual},'' \href{http://dx.doi.org/10.1007/JHEP05(2018)183}{{\em
  JHEP} {\bfseries 05} (2018) 183},
  \href{http://arxiv.org/abs/1711.08467}{{\ttfamily arXiv:1711.08467
  [hep-th]}}.

\bibitem{Almheiri:2014cka}
A.~Almheiri and J.~Polchinski, ``{Models of AdS$_{2}$ backreaction and
  holography},'' \href{http://dx.doi.org/10.1007/JHEP11(2015)014}{{\em JHEP}
  {\bfseries 11} (2015) 014}, \href{http://arxiv.org/abs/1402.6334}{{\ttfamily
  arXiv:1402.6334 [hep-th]}}.

\bibitem{Jensen:2016pah}
K.~Jensen, ``{Chaos in AdS$_2$ Holography},''
  \href{http://dx.doi.org/10.1103/PhysRevLett.117.111601}{{\em Phys. Rev.
  Lett.} {\bfseries 117} no.~11, (2016) 111601},
  \href{http://arxiv.org/abs/1605.06098}{{\ttfamily arXiv:1605.06098
  [hep-th]}}.

\bibitem{Engelsoy:2016xyb}
J.~Engels\"oy, T.~G. Mertens, and H.~Verlinde, ``{An investigation of AdS$_{2}$
  backreaction and holography},''
  \href{http://dx.doi.org/10.1007/JHEP07(2016)139}{{\em JHEP} {\bfseries 07}
  (2016) 139}, \href{http://arxiv.org/abs/1606.03438}{{\ttfamily
  arXiv:1606.03438 [hep-th]}}.

\bibitem{Cvetic:2016eiv}
M.~Cvetic and I.~Papadimitriou, ``{AdS$_{2}$ holographic dictionary},''
  \href{http://dx.doi.org/10.1007/JHEP12(2016)008}{{\em JHEP} {\bfseries 12}
  (2016) 008}, \href{http://arxiv.org/abs/1608.07018}{{\ttfamily
  arXiv:1608.07018 [hep-th]}}. [Erratum: JHEP 01, 120 (2017)].

\bibitem{Nayak:2018qej}
P.~Nayak, A.~Shukla, R.~M. Soni, S.~P. Trivedi, and V.~Vishal, ``{On the
  Dynamics of Near-Extremal Black Holes},''
  \href{http://dx.doi.org/10.1007/JHEP09(2018)048}{{\em JHEP} {\bfseries 09}
  (2018) 048}, \href{http://arxiv.org/abs/1802.09547}{{\ttfamily
  arXiv:1802.09547 [hep-th]}}.

\end{thebibliography}\endgroup

\end{document}